\documentclass[11pt]{article}
\usepackage{graphicx,amstext,amssymb}

\begin{document}

\author{S.V. Akkelin$^{1}$,  Y. Hama$^{2}$, Iu.A.
Karpenko$^{1}$, Yu.M. Sinyukov$^{1}$}
\title{Hydro-kinetic approach to relativistic heavy ion collisions}
 \maketitle

\begin{abstract}
We develop a combined hydro-kinetic approach which incorporates a
hydrodynamical expansion of the systems formed in
\textit{A}+\textit{A} collisions and their dynamical decoupling
described by escape probabilities. The method corresponds to a
generalized relaxation time ($\tau_{\text{rel}}$) approximation
for the Boltzmann equation applied to inhomogeneous expanding
systems; at small $\tau_{\text{rel}}$ it also allows one to catch
the viscous effects in hadronic component - hadron-resonance gas.
We demonstrate how the approximation of sudden freeze-out can be
obtained within this dynamical picture of continuous emission and
find that hypersurfaces, corresponding to a sharp freeze-out
limit, are momentum dependent. The pion $m_{T}$ spectra are
computed in the developed hydro-kinetic model, and compared with
those obtained from ideal hydrodynamics with the Cooper-Frye
isothermal prescription. Our results indicate that there does not
exist a universal freeze-out temperature for pions with different
momenta, and support an earlier decoupling of higher $p_{T}$
particles. By performing numerical simulations for various initial
conditions and equations of state we identify several
characteristic features of the bulk QCD matter evolution preferred
in view of the current analysis of heavy ion collisions at RHIC
energies.
\end{abstract}

\begin{center}

{\small \textit{$^{1}$ Bogolyubov Institute for Theoretical
Physics, Metrolohichna str. 14b, 03680 Kiev-143,  Ukraine  \\
[0pt] }} {\small \textit{$^{2}$ Instituto de F\'{\i}sica,
Universidade de S\~ao Paulo, S\~ao Paulo, SP, C.P. 66318,
05315-970,
Brazil \\[0pt] }}


\end{center}

\section{Introduction}

Hydrodynamic models successfully describe basic features of high
energy nuclear collisions at CERN SPS and especially at BNL RHIC
(for reviews see, e.g., Refs. \cite{Heinz,Hama1,Nonaka1}), where
the utilization of ideal hydrodynamics was supported by
theoretical results: it was advocated \cite{Shuryak} that
deconfined matter behaves like a perfect liquid. The hydrodynamic
approach to \textit{A}+\textit{A} collisions implies a very hot
and dense strongly interacting matter as the initial state. Such a
state is assumed to be formed soon after the collision, then
expands hydrodynamically until the stage when the picture of the
continuous medium is destroyed. Roughly, it happens when the mean
free path of particles becomes comparable with the smallest
characteristic dimension of the system: its geometrical size or
hydrodynamic length. This approach allows one to account for the
complicated evolution of the system at a preconfined stage and in
the vicinity of the possible phase transitions by means of a
corresponding equation of state (EoS). By studying the spectra of
different particle species versus the initial conditions  and EoS,
one could get information on the thermal partonic stage of the
``Little Bang" and further system evolution, which could be also
conclusive in the context of discriminating the possible phase
transitions.

The problem is, however, whether the predicted momentum spectra,
obtained in hydrodynamic models for given initial conditions and
EoS,  are unambiguous. These spectra depend not only on the
initial conditions but also on the final conditions of
hydrodynamic expansion since  one cannot use hydrodynamic
equations for infinitely large times because  the resulting very
small densities destroy the picture of continuous medium of real
particles. Evidently, the simplest  receipt of the spectrum
calculation is the Cooper-Frye prescription (CFp) \cite{Cooper}
which ignores the post-hydrodynamic (kinetic) stage of matter
evolution and assumes that perfect fluid hydrodynamics is valid
till some 3D hypersurface, e.g., as was supposed by Landau
\cite{Landau}, till the isotherm $T\simeq m_{\pi}$, where sudden
transition from local thermal equilibrium to free streaming is
assumed.

 It has been known for a long time that CFp leads to
inconsistencies \cite{Sin4,Bugaev}, if the freeze-out hypersurface
contains the non-space-like sectors, and should be modified to
exclude formally negative contributions to the particle number at
the corresponding momenta. The simplest prescription is to present
the distribution function as a product of a local thermal
distribution and the step function like $\theta
(p_{\mu}n^{\mu}(x))$ \cite{Bugaev}, $n^{\mu }n_{\mu}=\pm 1$, where
$n^{\mu}$ is a time-like or space-like outward normal to a
freeze-out hypersurface $\sigma $. Thereby freeze-out is
restricted to those particles for which $p_\mu n^\mu(x)$ is
positive. This receipt was used in Ref. \cite{Borysova} to
describe particle emission from enclosed freeze-out hypersurface
with non-space-like sectors and it was found that a satisfactory
description of central \textit{Au}+\textit{Au} collisions at RHIC
was reached for a physically reasonable set of parameters. The
main features of the experimental data were reproduced: in
particular, the obtained ratio of the outward to sideward
interferometry radii is less than unity and decreases with
increasing transverse momenta of pion pairs. Thereby, the results
of Ref. \cite{Borysova} clearly indicate that early particle
emission off the surface of the hydrodynamically expanding
fireball could be essential for proper description of matter
evolution in  \textit{A}+\textit{A} collisions.

However, sharp freeze-out at some 3D hypersurfaces is a rather
rough approximation of the spectrum formation, because  the
particle emission process of fireballs created in high energy
heavy ion collisions is gradual in time.  Results of many studies
based on cascade models contradict the idea of sudden freeze-out
and demonstrate that in fact particles are emitted from the  4D
volume during the whole period of the system evolution, and
deviations from local equilibrium conditioned by continuous
emission should take place (see, e.g., \cite{Bravina}). Moreover,
freeze-out hypersurfaces typically contain non-space-like parts
that lead to a problem with energy-momentum conservation law in
realistic dynamical models \cite{Bugaev}. This concerns also
hybrid models \cite{hybrid} where the transport model matches
hydrodynamics on such a  kind of isothermal hypersurface of
hadronization \cite{Bugaev1}.\footnote{ Note that in this hybrid
picture the initial conditions for hadronic cascade calculations
could be formulated also on some (arbitrary) {\it space-like}
hypersurface where, however, hadronic distributions are deviated
from the local equilibrium, in particular, because of an opacity
effect for hadrons which are created during a ``mixed" stage of
phase transition. These nonequilibrium effects could seriously
influence the results of hybrid models in its modern form
\cite{hybrid}.}

 An attempt to introduce 4D continuous emission in
hydrodynamic framework has been done in Ref. \cite{Hama2} within a
simple ideal hydrodynamics, and also supported by numerical
transport codes calculations, e.g., Ref. \cite{Molnar}. It was
found in these papers that the transport freeze-out process is
similar to evaporation: high-$p_{T}$ particles freeze out early
from the surface, while low-$p_{T}$ ones decouple later from the
system's center.

The idea of continuous decoupling  was further developed based on
Boltzmann equations in Ref. \cite{AkkSin1} where, in particular,
an approximate method that accounts for the  back reaction of the
emission on the fluid dynamics was proposed. It is worth noting
that the back reaction   is not reduced just to an energy-momentum
recoiling of emitted particles on the  expanding thermal medium,
but also leads to a rearrangement of the medium, producing a
deviation of its state from the local equilibrium, accompanied by
changing of the local temperature, densities and collective
velocity field. This complex effect is mainly a  consequence of
the impossibility  to split the evolution of the single finite
system of hadrons into the two components: expansion of the
interacting locally equilibrated medium and free streaming of
emitted  particles, which the system  consists of. Such a
splitting, accounting only for the momentum-energy conservation
law, contradicts  the underlying dynamical equations such as a
Boltzmann one \cite{AkkSin1}. In view of this, the ideas proposed
and results obtained in the quasiclassical approach should be a
clue for a quantum treatment  of the problem which recently begins
to to be developed \cite{Cramer}.\footnote{See also the  very
recent contribution to this field in Ref. \cite {Knoll}.} Note,
however, that the estimates of the influence of the quantum
effects, such as the distortion of the wave function, on the
spectra and Bose-Einstein correlations testify to be  relatively
small corrections to the quasiclassical approximation
\cite{Pratt}.

The aim of the present work is twofold. First, we develop the
formalism of the hydro-kinetic approach and propose approximations
for practical calculations. The problem of self-consistency of the
method accounting simultaneously for both the particle emission
and fluid (in general, viscous) dynamics is studied in detail.
Special attention is also paid to the discussion of the
applicability conditions of the Cooper-Frye prescription for sharp
freeze-out.

Second, we develop a simple hydro-kinetic model of hadronic
emission that describes the evolution and emission of  particles
from a hydrodynamically expanding system undergoing a phase
transition. For the sake of simplicity we consider here one type
of escaping particle (pions) only and do not take into account in
numerical calculations the back reaction of the emission on fluid
dynamics. We study within such an approach what type of initial
conditions, equation of state, etc., are preferred in view of
current analysis of heavy ion collisions at RHIC energies.

\section{Hydro-kinetic formalism for heavy ion collisions}

 It was proposed in Ref. \cite{AkkSin1} to describe the hadronic
momentum spectra in \textit{A}+\textit{A} collisions based on the
escape function of particles which are gradually liberated from
hydrodynamically expanding systems. The escape function,
introduced in \cite{Hama2}, is calculated within the Boltzmann
equations in a specific approximation based on a hydrodynamic
approach. It was shown that such a picture corresponds to a
relativistic kinetic equation with the relaxation time
approximation for the collision term, where the relaxation time
tends to infinity, $\tau_{\text{rel}}\rightarrow \infty$, when $t
\rightarrow \infty$, indicating a transition to the free streaming
regime. For one component system the equation has the form:
\begin{eqnarray}
 \frac{p^{\mu}}{p_0}\frac{\partial f(x,p)}{\partial x^{\mu }}
 =-\frac{f(x,p)-f^{\text{l eq}}(x,p)}{\tau_{\text{rel}}(x,p)}.
 \label{boltz-1}
\end{eqnarray}
Here $f(x,p)$ is the phase-space distribution function,
$f^{\text{l eq}}(x,p)$ is the local equilibrium distribution with
local velocities, temperatures, and chemical potentials that
should be found from Eq. (\ref{boltz-1}) and the initial
$f_0(x,p)$, and $\tau_{\text{rel}}(x,p)$ is the relaxation time
(inverse rate of collisions in gases),
\begin{eqnarray}
\tau_{\text{rel}}(x,p)=
\frac{p_{0}\tau^*_{\text{rel}} (x,p)}{p^{\mu}u_{\mu}}.
 \label{boltz-2}
\end{eqnarray}
Here $\tau^*_{\text{rel}} (x,p)$ is related to the local fluid
rest frame (local rest frame of the energy flow) where the
collective four-velocity is $u_{\mu}=(1,\mathbf{0})$. The
relaxation time depends on the cross section and is a functional
of $f^{\text{l eq}}(x,p)$.

As it is well known \cite{Huang,Groot}, such kinds of equations at
$\tau_{\text{rel}} \ll \tau_{\text{exp}}$ (inverse of expansion
rate) describe in the first approximation the viscosity effects in
gases with a coefficient of shear viscosity
$\eta\propto\tau_{\text{rel}}nT$. Therefore the method explained
below catches in the first approximation also the viscosity
effects in an expanding hadronic gas, characterized by fields of
temperatures $T$ and particle densities $n$. The viscosity effects
in the quark-gluon plasma (QGP) evolution cannot be described in
this way because  strongly interacting QGP is not a gas, but
almost an ideal liquid \cite{Shuryak}.

The formal solution of Eq. (\ref{boltz-1}) can be presented
in the following form:
\begin{eqnarray}
f(t,{\bf r},p) =f(t_{0},{\bf r}-\frac{{\bf p}}{p_{0}}
(t-t_{0}),p)\exp\left\{ -\stackrel{t}{%
%TCIMACRO{\underset{t_{0}}{\int }}%
%BeginExpansion
\mathrel{\mathop{\int }\limits_{t_{0}}}%
%EndExpansion
}\frac{1}{\tau_{\text{rel}}(s,{\bf r}-\frac{{\bf p}}{p_{0}}
(t-s),p)}ds\right\} +
\nonumber \\
\stackrel{t}{%
%TCIMACRO{\underset{t_{0}}{\int }}%
%BeginExpansion
\mathrel{\mathop{\int }\limits_{t_{0}}}%
%EndExpansion
}\frac{f^{\text{l eq}}(t^{\prime },{\bf r}-\frac{{\bf p}}{p_{0}}(t-t^{\prime }),p)}{%
\tau_{\text{rel}}(t^{\prime },{\bf r}-\frac{{\bf p}}{p_{0}}(t-t^{\prime }),p)}%
\exp \left\{ -\stackrel{t}{%
%TCIMACRO{\underset{t^{\prime }}{\int }}%
%BeginExpansion
\mathrel{\mathop{\int }\limits_{t^{\prime }}}%
%EndExpansion
}\frac{1}{\tau_{\text{rel}}(s,{\bf r}-\frac{{\bf
p}}{p_{0}}(t-s),p)}ds\right\} dt^{\prime},
 \label{boltz-3}
\end{eqnarray}
where $f(t_{0},{\bf r},p)$ is the initial distribution at
$t=t_{0}$. The relaxation time $\tau^*_{\text{rel}}$ as well as
the local equilibrium distribution function $f^{\text{l eq}}$ are
functionals of hydrodynamic variables: temperature $T$, chemical
potential $\mu$, and collective four-velocity $u_{\mu}$. The
space-time dependence of the corresponding variables are
determined by demanding the local conservation of the
energy-momentum with tensor $T^{\mu\nu}(x)$ and, if necessary, net
particle number, with current $n^{\mu}(x)$ (assuming no particle
production)

\begin{eqnarray}
\partial_{\mu}T^{\mu \nu}(x)=0,
\label{cons-1} \\
\partial_{\mu}n^{\mu}(x)=0,
\label{cons-2}
\end{eqnarray}
where (see, e.g., Ref. \cite{Groot})
\begin{eqnarray}
T^{\mu \nu}(x)= \int \frac{d^{3}k}{k_{0}}k^{\mu \nu}f(x,k),
\label{tens-def1} \\
n^{\mu}(x)=\int \frac{d^{3}k}{k_{0}}k^{\mu}f(x,k).
\label{number-def1}
\end{eqnarray}
These conservation laws lead to rather complicated equations for
hydrodynamic variables. It is worth noting that for an  expanding
system the relaxation time $\tau^*_{\text{rel}}(x,p)$ increases
with time and, therefore, the deviations from local equilibrium
increase too, thereby preventing a use of the widely applied
approximate methods based on the expansion of the distribution
function in the vicinity of the local equilibrium.

Then to solve the kinetic equation (\ref{boltz-1}), in accordance
with the conservation laws (\ref{cons-1}) and (\ref{cons-2}), we
need an approximate method that could be applied even for strong
deviations from local equilibrium. It is not our aim here to
suggest an exclusive solution of the problem. Rather some
arguments are presented below by the example of the relativistic
one component Boltzmann  gas with particle number conservation,
\begin{eqnarray}
f^{\text{l eq}}(x,p)=(2 \pi)^{-3}\exp \left
(-\frac{p^{\mu}u_{\mu}+ \mu }{T} \right ), \label{leq-1}
\end{eqnarray}
to show that such a method could be developed based on the
following procedure.

To take into account nonequilibrium effects accompanying the
particle emission in inhomogeneous violently expanding systems, we
utilize the integral representation (\ref{boltz-3}) of kinetic
equation (\ref{boltz-1}). Then, performing a partial integration
of the second term in Eq. (\ref{boltz-3}) and, assuming that
$f(t_{0},{\bf r},p)=f^{\text{l eq}}(t_{0},{\bf r},p)$, one can
decompose the distribution function to a local equilibrium part,
$f^{\text{l eq}}$, and a part describing a deviation from the
local equilibrium behavior, $g$:
\begin{eqnarray}
f=f^{\text{l eq}}(x,p)+g(x,p), \label{boltz-4}
\end{eqnarray}
where
\begin{eqnarray}
g(x,p)=-\stackrel{t}{%
%TCIMACRO{\underset{t_{0}}{\int }}%
%BeginExpansion
\mathrel{\mathop{\int }\limits_{t_{0}}}%
%EndExpansion
}\frac{df^{\text{l eq}}(t^{\prime },{\bf r}-\frac{{\bf p}}{p_{0}}(t-t^{\prime }),p)%
}{dt^{\prime }}\exp \left\{ -\stackrel{t}{%
%TCIMACRO{\underset{t^{\prime }}{\int }}%
%BeginExpansion
\mathrel{\mathop{\int }\limits_{t^{\prime }}}%
%EndExpansion
}\frac{1}{\tau_{\text{rel}}(s,{\bf r}-\frac{{\bf
p}}{p_{0}}(t-s),p)}ds\right\} dt^{\prime }. \label{boltz-5}
\end{eqnarray}
 Note that both functions, $f^{\text{l eq}}$ and $g$, are
functionals of hydrodynamic variables, $g$ depends also on the
relaxation time $\tau_{\text{rel}}$ that defines the mean time
interval between collisions, and $\tau_{\text{rel}}$ depends in
its turn on the distribution function $f^{\text{l eq}}$ and the
cross section. The evolution of the distribution function $f(x,p)$
should satisfy the energy-momentum conservation and, because
$T^{\mu\nu}[f]=T^{\mu\nu}[f^{\text{l eq}}+g]=T^{\mu\nu}
[f^{\text{l eq}}]+ T^{\mu\nu}[g]$ for systems where the
interaction energy can be neglected, it takes the form of
hydrodynamic equations for the perfect fluid with ``source",
\begin{eqnarray}
\partial_{\nu}T^{\nu
\beta}[f^{\text{l eq}}]=G^{\beta}[g], \label{sourse-1}
\end{eqnarray}
where
\begin{eqnarray}
G^{\beta}[g]=- \partial_{\nu} T^{\nu\beta}[g]. \label{sourse-2}
\end{eqnarray}
The equation that takes into account the conservation of
particle number has a similar form:
\begin{eqnarray}
\partial_{\nu}n^{\nu}[f^{\text{l eq}}]=S[g], \label{sourse-3}
\end{eqnarray}
where
\begin{eqnarray}
S[g]=- \partial_{\nu} n^{\nu }[g]. \label{sourse-4}
\end{eqnarray}

To find an approximate solution of Eqs.
(\ref{sourse-1})-(\ref{sourse-4}), one can solve the equations
\begin{eqnarray}
\partial_{\nu}T^{\nu
\mu}[f^{\text{l eq}}]= 0, \label{id-1} \\
\partial_{\nu}n^{\nu}[f^{\text{l eq}}]= 0, \label{id-2}
\end{eqnarray}
and, thereby, utilize the hydrodynamic variables in the perfect
fluid approximation. Namely, the hydrodynamic variables in this
approximation can be used to calculate the deviation from local
equilibrium $g(x,p)$ according to Eq. (\ref{boltz-5}) and, then,
``source" terms $G^{\beta}[g]$ and $S[g]$ on the right-hand sides
of Eqs. (\ref{sourse-1}) and (\ref{sourse-3}). Then the left-hand
sides of these equations are functionals of local equilibrium
functions and have the simple ideal fluid forms, while the
right-hand sides associated with a ``source" are explicit
functions which describe a deviation from the local equilibrium
and depend on hydrodynamic variables in the perfect fluid
approximation:
\begin{eqnarray}
\partial_{\nu}T^{\nu
\beta}[f^{\text{l
eq}}(T,u_{\mu},\mu)]=G^{\beta}[T_{\text{id}},u^{\text{id}}_{\mu},\mu_{\text{id}},\tau_{\text{rel}}^{\text{id}}],
\label{sourse-5} \\
\partial_{\nu}n^{\nu}[f^{\text{l eq}}(T,u_{\mu},\mu)]=S[T_{\text{id}},u^{\text{id}}_{\mu},\mu_{\text{id}},\tau_{\text{rel}}^{\text{id}}],
\label{sourse-6}
\end{eqnarray}
where, for one component Boltzmann gas with elastic collisions
only, the relaxation time $\tau_{\text{rel}}^{\text{id}}$ is the
inverse of collision rate in ideal fluid, $R^{\text{id}}(x,p)$,
and has the following form (in the co-moving frame):
\begin{eqnarray}
\frac{1}{\tau_{\text{rel}}^{\text{id}*}(x,p)} =
R^{\text{id}}(x,p)= \int \frac{d^3k}{(2\pi)^3}
\exp\left(-\frac{E_k - \mu_{id}(x)}{T_{id}(x)}\right ) \sigma(s)
\frac{\sqrt{s(s-4m^2)}}{2 E_{p}E_{k}}. \label{boltz-7}
\end{eqnarray}
Here $E_p=\sqrt{\mathbf{p}^2+m^2}$, $E_k=\sqrt{\mathbf{k}^2+m^2}$,
$s=(p+k)^2$ is the squared c.m. energy of the pair, and
$\sigma(s)$ is the corresponding cross section. A solution
$(T(x),u_\mu(x),\mu(x))$ of Eqs. (\ref{sourse-5}),
(\ref{sourse-6}) accounts for the back reaction of the emission
process on hydro-evolution and provides us with the hydrodynamic
parameters which finally should be used to calculate the locally
equilibrated part $f^{\text{l eq}}(x,p)$ of the complete
distribution function $f(x,p)$. Then, the distribution function
obtained in this way, $f(x,p)=f^{\text{l eq}} [T,u_\mu,\mu]+
g[T_{\text{id}},u^{\text{id}}_\mu,
\mu_{\text{id}},\tau_{\text{rel}}^{\text{id}}]$, satisfies the
conservation laws, takes into account the nonequilibrium
peculiarities of the evolution and is constructed in agreement
with the corresponding EoS. Of course, this scheme allows us to
make the next iterations in solving Eq. (\ref{boltz-1}). Note also
that because  the ``source" term on the  right-hand side  of Eq.
(\ref{sourse-5}) is a known function, the causality is preserved
in this description of dissipative systems.

\section{Kinetics of the freeze-out in hydro-kinetic approach}

 The approach developed in the previous section allows us to
study an important problem: under which conditions the
Landau/Cooper-Frye prescription (CFp) of sudden freeze-out is
valid and how, then, to define the corresponding 3D freeze-out
hypersurface. The CFp is traditionally utilized when hydrodynamics
is applied to describe the later stage of the matter evolution in
\textit{A}+\textit{A} collisions, and also is a basic ingredient
of various hydro-motivated parametrizations (see, e.g.,
\cite{Borysova,blast}). A widely used heuristic freeze-out
criterion for finding the freeze-out hypersurface of a violently
expanding system is  either the comparability of the hydrodynamic
rate of expansion with the kinetic rate of collisions
\cite{Bondorf} or of the mean free path of particles with the
geometrical size of the system \cite{Landau,Navarra}.  While the
problem was extensively studied before (see, e.g., Refs.
\cite{Sin4,Bugaev,AkkSin1,Grassi}), a complete understanding of
the conditions allowing  the utilization of  the sharp freeze-out
approximation (that corresponds formally to sudden transition from
local equilibrium to free streaming) of spectra formation  and an
unambiguous definition of the corresponding freeze-out
hypersurface are still absent. A discussion in this section based
on the analytic approximation to a calculation of momentum spectra
could be, in our opinion, useful for understanding the conditions
of applicability of the CFp and improvement of it, if necessary.

Let us integrate over space variables distribution (\ref{boltz-3})
to represent the particle momentum density at large enough time,
$t\rightarrow\infty$, when particles in the system stop to
interact:
\begin{eqnarray}
\frac{d^3N}{d^3p}(t)\equiv n(t,p)=\int d^{3}r f(t, \mathbf{r}, p).
\label{mom-sp1}
\end{eqnarray}
The result can be presented in the general form found in Ref.
\cite{AkkSin1}:
\begin{eqnarray}
n(t\rightarrow \infty,p) = \int d^{3}r f(t_0,{\bf r},p){\cal
P}(t_0,{\bf r},p) +
\nonumber \\
 \int d^{3}r\stackrel{t}{%
%TCIMACRO{\underset{t_{0}}{\int }}%
%BeginExpansion
\mathrel{\mathop{\int }\limits_{t_{0}}}%
%EndExpansion
} dt^{\prime }F^{\text{gain}}(t^{\prime },{\bf r},p){\cal
P}(t^{\prime},{\bf r},p).
 \label{mom-sp-base}
\end{eqnarray}
The term $F^{\text{gain}}(x,p)$ corresponds to ``gain" term in the
Boltzmann equation and is associated with  the number of particles
which came to the phase-space point $(x,p)$ just after the
interaction in the system.  The probability ${\cal P}(x,p)$ for a
particle to escape the system from space-time point
$x=(t^{\prime},{\bf r})$ is expressed explicitly in terms of the
rate of collisions $R$ along the world line of the free particle
with momentum $\textbf{p}$,
\begin{equation}
{\cal P}(t^{\prime},{\bf
r},p)=\exp\left(-\int\limits_{t^{\prime}}^{\infty}ds R(s,{\bf
r}+\frac{{\bf p}}{p_{0}}(s-t^{\prime}),p)\right).
\label{prob-calc}
\end{equation}
 The first term in Eq. (\ref{mom-sp-base}) describes the
contribution to the momentum spectrum from particles that are
emitted from the very initial time, while the second one describes
the continuous emission with emission density
$S=F^{\text{gain}}{\cal P}$ from 4D volume delimited by the
initial and final (where particles stop to interact) 3D
hypersurfaces \cite{AkkSin1}. In a particular case, which we
consider in this article, $R(x,p)=1/\tau_{\text{rel}}(x,p)$ and
$F^{\text{gain}}(x,p)=f^{\text{l
eq}}(x,p)/\tau_{\text{rel}}(x,p)$.

To provide straightforward calculations leading  to the
Cooper-Frye approximation let us shift the  spacial variables,
${\bf r}^{\prime}={\bf r}+\frac{{\bf p}}{p_{0}}(t_0-t^{\prime})$,
in the second term of Eq. (\ref{mom-sp-base}) aiming to eliminate
the variable $t^{\prime}$ in the argument of the function
$R=1/\tau_{\text{rel}}$ in (\ref{prob-calc}). Then
\begin{eqnarray}
n(t,p) = \int d^{3}r f(t_0,{\bf r},p)
\exp\left\{-\stackrel{t}{%
%TCIMACRO{\underset{t_{0}}{\int }}%
%BeginExpansion
\mathrel{\mathop{\int }\limits_{t_{0}}}%
%EndExpansion
}\frac{1}{\tau_{\text{rel}}(s,{\bf r}+\frac{{\bf
p}}{p_{0}}(s-t_{0}),p)}ds\right\} +
\nonumber \\
 \int d^{3}r^{\prime} \stackrel{t}{%
%TCIMACRO{\underset{t_{0}}{\int }}%
%BeginExpansion
\mathrel{\mathop{\int }\limits_{t_{0}}}%
%EndExpansion
} dt^{\prime } f^{\text{l eq}}(t^{\prime },{\bf r}^{\prime
}+\frac{{\bf p}}{p_{0}}(t^{\prime }-t_{0}),p)Q (
t^{\prime},\textbf{r}^{\prime},p),
 \label{mom-sp2}
\end{eqnarray}
 where
\begin{eqnarray}
Q(t^{\prime},\textbf{r}^{\prime},p)\equiv\frac{1}{%
\tau_{\text{rel}}(t^{\prime },{\bf r}^{\prime }+\frac{{\bf p}}{p_{0}}(t^{\prime }-t_{0}),p)}%
\exp \left\{ -\stackrel{t}{%
%TCIMACRO{\underset{t^{\prime }}{\int }}%
%BeginExpansion
\mathrel{\mathop{\int }\limits_{t^{\prime }}}%
%EndExpansion
}\frac{1}{\tau_{\text{rel}}(s,{\bf r}^{\prime  }+\frac{{\bf
p}}{p_{0}}(s-t_{0 }),p)}ds\right\}. \label{mom-sp3}
\end{eqnarray}
 Here, $f^{\text{l eq}}$, $\tau_{\text{rel}}$ are functionals
of temperature, hydrodynamic four-velocity, and chemical
potential, and these quantities are governed by hydrodynamic
equations (\ref{sourse-1})-(\ref{sourse-4}). Note that
\begin{eqnarray}
Q ( t^{\prime},\textbf{r}^{\prime},p) = \frac{d}{dt^{\prime }}
P(t^{\prime},\textbf{r}^{\prime},p),
 \label{mom-spectr-1}
\end{eqnarray}
where $P(t^{\prime},\textbf{r}^{\prime},p)$ is connected with the
escape probability ${\cal P}(t^{\prime},{\bf r},p)$:
\begin{eqnarray}
P(t^{\prime},\textbf{r}^{\prime},p) \equiv
\exp \left\{ -\stackrel{t}{%
%TCIMACRO{\underset{t^{\prime }}{\int }}%
%BeginExpansion
\mathrel{\mathop{\int }\limits_{t^{\prime }}}%
%EndExpansion
}\frac{1}{\tau_{\text{rel}}(s,{\bf r}^{\prime  }+\frac{{\bf
p}}{p_{0}}(s-t_{0 }),p)}ds\right\}= {\cal P}(t^{\prime},{\bf
r}^{\prime  }+\frac{{\bf p}}{p_{0}}(t^{\prime}-t_{0 }),p).
 \label{mom-spectr-2}
\end{eqnarray}
Therefore
\begin{equation}
\stackrel{\infty}{%
%TCIMACRO{\underset{t^{\prime }}{\int }}%
%BeginExpansion
\mathrel{\mathop{\int
}\limits_{t_0}}}dt^{\prime}Q(t^{\prime},\textbf{r}^{\prime},p)=1-{\cal
P}(t_0,\textbf{r}^{\prime},p). \label{norm}
\end{equation}
 The 4D emission density has the form
\begin{eqnarray}
S( t^{\prime},\textbf{r}^{\prime},p)=f^{\text{l eq}}(t^{\prime
},{\bf r}^{\prime }+\frac{{\bf p}}{p_{0}}(t^{\prime }-t_{0}),p)Q (
t^{\prime},\textbf{r}^{\prime},p). \label{emis-def}
\end{eqnarray}

In order to have a tractable approach and set up the   conditions
of validity of  CFp,
  let us assume that at each $\textbf{r}^{\prime}$ and $\textbf{p}$ there is a maximum in $t^{\prime
 }$ of the emission function $S( t^{\prime},\textbf{r}^{\prime},p)$ inside the interval
 $[t_{0},t]$, and that the position of the  maximum,
$t^{\prime }=t^{\prime }_{\sigma}(\textbf{r}^{\prime},p)$,  is
mainly conditioned by $Q ( t^{\prime},\textbf{r}^{\prime},p)$.
Then corresponding  hypersurface $t_{\sigma}^{\prime
}(\mathbf{r}^{\prime },p)$ is defined by the conditions
\begin{eqnarray}
\frac{d Q ( t^{\prime},\textbf{r}^{\prime},p)}{d t^{\prime
}}|_{t^{\prime }=t_{\sigma}^{\prime}}=0,
 \label{saddle-0-1} \\
\frac{d^{2} Q ( t^{\prime},\textbf{r}^{\prime},p)}{d t^{\prime
 2}}|_{t^{\prime }=t_{\sigma}^{\prime}}<0, \label{saddle-0-2}
\end{eqnarray}
and, utilizing Eq. (\ref{mom-sp3}), we get
\begin{eqnarray}
\frac{d \tau_{\text{rel}}(t^{\prime },{\bf r}^{\prime }+\frac{{\bf p}}{p_{0}}(t^{\prime }-t_{0}),p)%
|_{t^{\prime }=t_{\sigma}^{\prime}}}{dt^{\prime }}=1,  \label{saddle-1} \\
\frac{d^{2}\tau_{\text{rel}}(t^{\prime },{\bf r}^{\prime }+\frac{{\bf p}}{p_{0}}(t^{\prime }-t_{0}),p)%
|_{t^{\prime }=t_{\sigma}^{\prime}}}{dt^{\prime 2}}>0,
\label{saddle-2}
\end{eqnarray}
where $d/d t^{\prime }$ is the full time derivative.

Let us demonstrate that Eqs. (\ref{saddle-1}),
(\ref{saddle-2}) generalize the heuristic freeze-out  criterion \cite{Bondorf}.
According to the latter, the freeze-out happens when
\begin{eqnarray}
\tau_{\text{scat}}\approx\tau_{\text{exp}}\,, \label{crit-1}
\end{eqnarray}
where $\tau_{\text{scat}}(x)$ is the mean time interval between successive scattering events,
\begin{eqnarray}
\tau_{\text{scat}}(x)=\frac{1}{\langle v \sigma \rangle (x) n(x)}\,,
\label{crit-2}
\end{eqnarray}
$v$ is the relative velocity between the scattering particles,
$n(x)$ is the particle number density and $\sigma$ is the
corresponding total cross section and the sharp brackets mean an
average over the local thermal distribution. The inverse
hydrodynamic expansion rate $\tau_{\text{exp}}(x)$ is the
collective expansion time scale,
\begin{eqnarray}
\tau_{\text{exp}}(x) = - \left (
\frac{1}{n(x)}u^{\mu}\partial_{\mu}n \right )^{-1}= - \left (
\frac{1}{n(x)}\frac{\partial n}{\partial t^{*}}\right )^{-1},
\label{crit-3}
\end{eqnarray}
where $t^{*}$ is the proper time in the local rest frame of
the fluid.

To demonstrate that the  heuristic freeze-out criterion
(\ref{crit-1}) follows from Eqs. (\ref{saddle-1}) and
(\ref{saddle-2}) under certain conditions, we note first that
$\tau_{\text{scat}}(x)$ is similar to
$\tau_{\text{rel}}^{\text{id}*}$, see Eq. (\ref{boltz-7}). Then,
rewriting Eq. (\ref{saddle-1}) in the form
\begin{eqnarray}
\left ( \frac{\partial}{\partial
 t^{\prime }} + \frac{\mathbf{p}}{p_{0}} \frac{\partial}{\partial \mathbf{r}}\right )
\tau_{\text{rel}} (t^{\prime },{\bf r},p) |_{t^{\prime
}=t_{\sigma}^{\prime}, {\bf r}={\bf r}^{\prime}+\frac{{\bf
p}}{p_{0}}(t_{\sigma}^{\prime}-t_{0})}=1, \label{6-2nn}
\end{eqnarray}
we get from the above equation that
\begin{eqnarray}
- \left (\frac{1}{\tau_{\text{rel}} (t_{\sigma}^{\prime},{\bf
r},p)}\right )^{-1}\left ( \frac{\partial}{\partial
 t^{\prime }} + \frac{\mathbf{p}}{p_{0}} \frac{\partial}{\partial \mathbf{r}}\right ) \left (
\frac{1}{\tau_{\text{rel}}(t^{\prime },{\bf r},p)%
} \right )|_{t^{\prime }=t_{\sigma}^{\prime}, {\bf r}={\bf
r}^{\prime}+\frac{{\bf
p}}{p_{0}}(t_{\sigma}^{\prime}-t_{0})}= \nonumber \\
\frac{1}{\tau_{\text{rel}}(t_{\sigma}^{\prime},{\bf
r},p)}|_{t^{\prime}=t_{\sigma}^\prime,{\bf r}={\bf r}^\prime
+\frac{{\bf p}}{p_{0}}(t_{\sigma}^{\prime}-t_{0})}\,.
\label{6-8}
\end{eqnarray}
Now, if we turn to the local rest frame of fluid and neglect there
the momentum dependence of the above equation, as well as
deviations from local equilibrium, we get, accounting for Eqs.
(\ref{boltz-7}) and (\ref{crit-2}), that the left-hand side of Eq.
(\ref{6-8}) is approximately $-\left(\frac{1}{n}\frac{\partial
n}{\partial t^{\prime*}}\right)|_{t^{\prime
*}=t_{\sigma}^{\prime*}}$ while the right-hand side is
$1/\tau_{\text{rel}}^{\text{id}*}(t_\sigma^{\prime*}, {\bf
r}^{*})\approx 1/\tau_{\text{scat}}(t_\sigma^{\prime*}, {\bf
r}^{*})$, recovering thereby the criterion (\ref{crit-1}) of the
freeze-out. One can note, however, that unlike the heuristic
definition (\ref{crit-1}) the true freeze-out hypersurface
$t_{\sigma}^{\prime}({\bf r}^{\prime},p)$ depends on
momentum\footnote{The momentum dependence of freeze-out
hypersurface is conditioned by both momentum and spacial
dependencies of $\tau_{\text{rel}}$. Note that the momentum
dependence of the relaxation time was explicitly demonstrated
recently in Ref. \cite{Greiner} based on the numerical results of
parton cascade simulations.} and thereby particles of different
momenta freeze out on different hypersurfaces. This result has
already been observed in \cite{Hama2} in terms of escape
probabilities.

Now, we proceed to the Cooper-Frye representation of the
freeze-out process. For this aim it is convenient to introduce a
new variable in the second term of Eq. (\ref{mom-sp2}), namely,
\begin{eqnarray}
{\bf r} = {\bf r}^{\prime }+\frac{{\bf p}}{p_{0}}(t^{\prime
}_{\sigma }({\bf r}^{\prime },p)-t_{0})\,.  \label{mom-sp5}
\end{eqnarray}
Then
\begin{eqnarray}
{\bf r}^{\prime }={\bf r} - \frac{{\bf p}}{p_{0}}(t_{\sigma }({\bf
r},p)-t_{0}),  \label{mom-sp6}
\end{eqnarray}
and the corresponding Jacobian is
\begin{eqnarray}
 \det \left [ \frac{\partial
r^{\prime }_{i}}{\partial r_{j}} \right ]= 1-
\frac{\mathbf{p}}{p_{0}}\frac{\partial t_{\sigma }}{\partial {\bf
r}}. \label{c-f5}
\end{eqnarray}
This function is positive as will be shown below. Then Eq.
(\ref{mom-sp2}) takes the form
\begin{eqnarray}
n(t\rightarrow \infty,p) = \int d^{3}r f(t_0,{\bf r},p){\cal
P}(t_0,{\bf r},p)
 +
\nonumber \\
 \int d^{3}r \left (1 -
\frac{\mathbf{p}}{p_{0}}\frac{\partial t_{\sigma }}{\partial {\bf
r}}\right ) \stackrel{t}{%
%TCIMACRO{\underset{t_{0}}{\int }}%
%BeginExpansion
\mathrel{\mathop{\int }\limits_{t_{0}}}%
%EndExpansion
} dt^{\prime } f^{\text{l eq}}(t^{\prime },{\bf r}+\frac{{\bf
p}}{p_{0}}(t^{\prime }-t_{\sigma}),p)Q ( t^{\prime},{\bf r} -
\frac{{\bf p}}{p_{0}}(t_{\sigma }({\bf r},p)-t_{0}),p).
 \label{mom-sp10}
\end{eqnarray}
To estimate the accuracy of the Cooper-Frye approximation let us
apply the saddle-point method for the calculation of the second
term in Eq. (\ref{mom-sp10}) by the use of expansion of the
function
 $\ln Q ( t^{\prime},{\bf r} - \frac{{\bf
p}}{p_{0}}(t_{\sigma }({\bf r},p)-t_{0}),p)=
  \ln [\tau^{-1}_{\text{rel}}(t^{\prime},{\bf r}+\frac{{\bf
p}}{p^0}(t^{\prime}-t_{\sigma}({\bf r},p),p){\cal
P}(t^{\prime},{\bf r}+\frac{{\bf
p}}{p^0}(t^{\prime}-t_{\sigma}({\bf r},p),p)]$ (see Eq.
(\ref{mom-sp3})) near the point of maximum in $t^{\prime}$:
\begin{eqnarray}
Q ( t^{\prime},{\bf r} -
\frac{{\bf p}}{p_{0}}(t_{\sigma }({\bf r},p)-t_{0}),p)\approx \frac{\exp (-\stackrel{\infty}{%
%TCIMACRO{\underset{t_{0}}{\int }}%
%BeginExpansion
\mathrel{\mathop{\int }\limits_{t_{\sigma}}}%
%EndExpansion
}ds\frac{1}{\tau_{\text{rel}}(s,{\bf r}+\frac{{\bf
p}}{p_0}(s-t_{\sigma}({\bf
r},p))})}{\tau_{\text{rel}}(t_{\sigma}({\bf r},p),{\bf
r},p)}\exp(-(t^{\prime }-t_{\sigma}({\bf r},p))^{2}/2D^2({\bf
r},p)),
 \label{mom-sp4}
\end{eqnarray}
 where
\begin{eqnarray}
D^2({\bf r}, p)=\frac{\tau_{\text{rel}}(t_{\sigma}({\bf r},p),{\bf
r},p)}{\frac{d^{2}\tau_{\text{rel}}(t^{\prime },{\bf r}^{\prime
}+\frac{{\bf p}}{p_{0}}(t^{\prime }-t_{0}),p)%
}{dt^{\prime 2}}|_{t^{\prime }=t_{\sigma}({\bf r},p),{\bf
r}^{\prime }={\bf r} - \frac{{\bf p}}{p_{0}}(t_{\sigma }({\bf
r},p)-t_{0})} }. \label{D}
\end{eqnarray}

Let us suppose that probability ${\cal P}(t_0,{\bf r},p)$ for
particles to escape just at the initial moment  is negligible,
except, of course, the periphery of the system at $t=t_0$ which,
however, gives a relatively small contribution to the spectra.
Then, taking into account that in the region of freeze-out the
particle suffers
the last collision and, thus, $\stackrel{\infty}{%
%TCIMACRO{\underset{t_{0}}{\int }}%
%BeginExpansion
\mathrel{\mathop{\int }\limits_{t_{\sigma}}}%
%EndExpansion
}ds\tau^{-1}_{\text{rel}}(s,{\bf r}+\frac{{\bf
p}}{p^0}(s-t_{\sigma}({\bf r},p),p)\simeq 1$
($\tau^{-1}_{\text{rel}}$ is the rate of collisions), and also
accounting for the normalizing condition (\ref{norm}) for $Q$, one
can integrate Eq. (\ref{mom-sp4}) over $t'$ and estimate the
temporal width of the emission zone
\begin{equation}
D({\bf
r},p)=\frac{1}{\sqrt{2\pi}}\tau_{\text{rel}}(t_{\sigma},{\bf
r},p)\exp \left (\stackrel{\infty}{%
%TCIMACRO{\underset{t_{0}}{\int }}%
%BeginExpansion
\mathrel{\mathop{\int }\limits_{t_{\sigma}}}%
%EndExpansion
}ds\tau^{-1}_{\text{rel}}(s,{\bf r}+\frac{{\bf
p}}{p^0}(s-t_{\sigma}({\bf r},p),p)\right )\simeq
\tau_{\text{rel}}(t_{\sigma},{\bf r},p). \label{width}
\end{equation}
The above relation means that at the freeze-out hypersurface
$\tau_{\text{rel}}\tau''_{\text{rel}} \simeq 1$, see Eq.
(\ref{D}).

Therefore if the temporal homogeneity length $\lambda({\bf r},p)$
of the distribution function $f^{\text{l eq}}$ near the four-point
$(t_{\sigma}({\bf r},p),{\bf r})$ is much larger then the width of
the emission zone, $\lambda({\bf r},p)\gg \tau_{\text{rel}}({\bf
r},p)$, then one can approximate $f^{\text{l eq}}(t^{\prime },{\bf
r}+\frac{{\bf p}}{p_{0}}(t^{\prime }-t_{\sigma}),p)$ by
$f^{\text{l eq}}(t_{\sigma },{\bf r},p)$ in Eq. (\ref{mom-sp10})
and perform integration over $t'$  accounting for normalizing
condition (\ref{norm}). As a result we get   momentum spectra in
the Cooper-Frye form:
\begin{eqnarray}
 p^0 n(t \rightarrow \infty,p) =
\mathrel{\mathop{\int }\limits_{\sigma(p)}}
d\sigma_{\mu}p^{\mu}f^{\text{l eq}}(x,p), \label{mom-sp14}
\end{eqnarray}
where
\begin{eqnarray}
d\sigma_{\mu}p^{\mu}= d^{3}r \left (p_{0} -
\mathbf{p}\frac{\partial t_{\sigma }}{\partial {\bf r}}\right ).
\label{mom-sp14-sigma}
\end{eqnarray}

 Now let us summarize the conditions for the utilization
of the Landau/Cooper-Frye approximation of sudden freeze-out. They
are the following:
\smallskip

(i) For each momentum $\textbf{p}$, there is a region of ${\bf r}$
where the emission function as well as the function $Q$ (see Eq.
(\ref{mom-sp3})) have a sharp maximum with temporal width $D({\bf
r},p)$. The formal conditions of the maximum are defined by Eqs.
(\ref{saddle-0-1}), (\ref{saddle-0-2}).
\smallskip

(ii) The width of the maximum, which in the case of one component
system is just the relaxation time (inverse of collision rate),
should be smaller than the corresponding temporal homogeneity
length of the distribution function: $\lambda({\bf r},p)\gg D({\bf
r},p) \simeq \tau_{\text{rel}}({\bf r},p)$.
\smallskip

(iii) The contribution to the spectra from the residual region of
${\bf r}$ where the saddle point method (Gaussian approximation
(\ref{mom-sp4}) and/or condition $\tau_{\text{rel}} \ll \lambda$)
is violated does not affect essentially the particle momentum
density. In other words, the space regions where for fixed
momentum $\textbf{p}$ either there is no  sharp maximum of the
emission function or phase-space density changes very rapidly are
not important for spectra formation. Note, that this condition is
most questionable and has to be checked in realistic dynamical
approaches for freeze-out, like hydro-kinetics, or transport
models with appropriate initial conditions for hadronic phase.
\smallskip

(iv) The escape probabilities ${\cal P}(t_0,{\bf r},p)$ for
particles to be liberated just from the initial hypersurface $t_0$
are small almost in the whole spacial region (except maybe
peripheral points) and so, one can neglect the first integral in
Eq. (\ref{mom-sp10}).
\smallskip

If the conditions (i) - (iv) are satisfied, then the momentum
spectra can be presented in Cooper-Frye form {\it in spite of the
fact that there is no  sudden freeze-out and the decaying region
has a finite temporal width} $D({\bf r},p) \simeq
\tau_{rel}(t_{\sigma}({\bf r},p),{\bf r},p) $. Also, what is very
important, such a generalized Cooper-Frye representation is
related to {\it the freeze-out hypersurface that depends on
momentum} $\textbf{p}$ and {\it does not necessarily enclose the
initially dense matter } (it will be demonstrated in the next
section by numerical calculations of emission functions).

In as much as the hypersurface $t_\sigma({\bf r},p)$ corresponds
to the maximum of the emission of the particles with momentum
$\textbf{p}$ into vacuum, these particles cannot be emitted from
point $(t_{\sigma}({\bf r},p),{\bf r})$ inward of the system. In
the latter case neither the emission function $S$ nor function $Q$
can have a maximum at this phase-space point; in fact, their
values are near zero. Even formally, in the Gaussian approximation
(\ref{mom-sp4}) for $Q$, validated in the
region of its maximal value, the integral $\stackrel{\infty}{%
%TCIMACRO{\underset{t_{0}}{\int }}%
%BeginExpansion
\mathrel{\mathop{\int }\limits_{t_{\sigma}}}%
%EndExpansion
}ds\tau^{-1}_{\text{rel}}(s,{\bf r}+\frac{{\bf
p}}{p^0}(s-t_{\sigma}({\bf r},p),p)\gg 1$, if the particle world
line crosses almost the whole system. The latter results in
$Q\rightarrow0$ and, therefore, completely destroys the
saddle-point  approximation (\ref{mom-sp4}) for  $Q$. If the
particle crosses some non-space-like part of the hypersurface
$\sigma$ moving inward of the system, it corresponds to the
condition $p^{\mu}d\sigma_{\mu}< 0$ \cite{Sin4}. Hence {\it
always} the value $p^{\mu}d\sigma_{\mu}(p)$ in the generalized
Cooper-Frye formula (\ref{mom-sp14}) is positive:
$p^{\mu}d\sigma_{\mu}(p)>0$ across the hypersurface where a fairly
sharp maximum of the  emission of particles with momentum
$\textbf{p}$ is situated;  and so the requirement
$p^{\mu}d\sigma_{\mu}(p)>0$ is a necessary condition for
$t_{\sigma}({\bf r},p)$ to be a true hypersurface of the maximal
emission. It means that hypersurfaces of maximal emission for a
given momentum $\textbf{p}$ may be open in space-time, not
enclosing the high-density matter at initial time $t_0$, and
different for different $\textbf{p}$. In other words, the momentum
dependence of the freeze-out hypersurface, defined by Eqs.
(\ref{saddle-1}) and (\ref{saddle-2}), naturally restricts the
freeze-out to those particles for which
$p^{\mu}d\sigma_{\mu}(\textbf{r},p)$ is positive.

Therefore, there are no negative contributions to the particle
momentum density from non-space-like sectors of the freeze-out
hypersurface, that is a well-known shortcoming of the Cooper-Frye
prescription \cite{Sin4,Bugaev}; the negative contributions could
appear only as a result of utilization of improper freeze-out
hypersurface that roughly ignores its momentum dependence and so
is common for all $\textbf{p}$. If, anyhow, such a common
hypersurface will be used, e.g., as the hypersurface of the
maximal particle number emission (integrated over $\textbf{p}$),
there is no possibility to justify the approximate expression for
momentum spectra similar to Eq. (\ref{mom-sp14}). Also, in that
case there is no common phenomenological prescription, based on
Heaviside step functions $\theta(p_{\mu}d\sigma_{\mu})$, which
allows one to eliminate the negative contributions to the momentum
spectra when $p^{\mu}d\sigma_{\mu}< 0$. The prescription, proposed
in \cite{Sin4}, eliminates the negative contributions in the way
which preserves the number of  particles in the fluid element
crossing the freeze-out hypersurface related to the maximum of
total (averaged over $\textbf{p}$) particle emission. Therefore it
takes into account that at the final stage the system is the only
holder of emitted particles. Another prescription \cite{Bugaev}
ignores the particle number conservation considering decaying
hadronic system rather as a star - practically unlimited reservoir
of emitted photons/particles. Both prescriptions have a problem
with momentum-energy conservation laws at freeze-out. The approach
developed here overcomes all above-mentioned problems by
considering the continuous dynamical freeze-out that is consistent
with Boltzmann equations and conservation laws.

One can see from Eq. (\ref{mom-sp4}) that in the general case with
nonzero thickness of the emission layer, $D \simeq
\tau_{\text{rel}} \neq 0$, the emission density $S$
(\ref{emis-def}) cannot be approximated by means of the local
equilibrium distribution function $f^{\text{l eq}}$ smeared by a
(proper) time factor $\exp(-(\tau-\tau_{0})^2/\delta\tau^{2})$
with constant $\tau_{0}$ and $\delta\tau^{2}$. Such an ansatz
cannot be used in place of the proper emission function as it is
often utilized in hydro-inspired parametrizations (see, e.g., Ref.
\cite{blast}) to take into account the gradual character of the
freeze-out process in heavy ion collisions. In  fact, parameters
$\tau_0$ and $\delta\tau^2$ should be dependent at least on the
 spacial coordinates, as it is explained in detail in Refs.
\cite{Borysova,AkkSin1}. Equations (\ref{mom-sp4}) and (\ref{D})
demonstrate that they depend on particle momentum as well and that
the emission is not locally isotropic.

\section{Pionic emission in specific hydro-kinetic models}

 In this section we present and discuss the results of
numerical calculations obtained in a few concrete realizations of
the hydro-kinetic model (HKM). For the sake of simplicity we
consider the emission of only one particle species (negative
pions, $\pi^{-}$) from the expanding fireball. Also we employ here
the ideal fluid approximation for hydrodynamic variables in the
integral representation (\ref{boltz-3}) to calculate the emission
densities and momentum spectra of pions. This approximation
results in a little (less than $10\%$ for our calculations)
violation of the conservation  laws. We describe, in all
simulations, the locally equilibrated state of hadrons by
relativistic Boltzmann distributions.

The numerical results presented in this section were obtained on
the basis of our original numerical 3D ideal hydro-code that was
developed based on the relativistic Godunov-type HLLE algorithm,
described in detail in Ref. \cite{rischke}. We use Bjorken
coordinates \cite{Bjorken} $\tau = \sqrt{t^{2}-z^{2}}$,
$\eta=\tanh^{-1}(z/t)$, instead of Cartesian ones, the
corresponding transformation of hydrodynamic equations and
conservative variables are described, e.g., in Refs.
\cite{Hama1,Hirano}. A second order of accuracy in time of this
algorithm is achieved by using the predictor-corrector scheme. To
achieve second order of accuracy in space we apply a linear
distribution of conservative variables inside each fluid cell.
Having determined the evolution in the ideal fluid approximation,
we proceed in calculating the emission densities using numerical
(tabulated) space-time dependencies of collective velocities and
thermodynamic quantities. An 8- and 16-point Gaussian quadrature
method implemented in ROOT \cite{ROOT} are used to calculate the
escape probability for each space-time position and particle
momentum. To calculate the resulting 4D integrals for particle
spectra we use Monte Carlo method and check convergence of the
results with increasing number of sampling points. We use ROOT for
the results output also.

   First, we present our results on a specific toy-HKM that
describes expanding one component relativistic ideal Boltzmann gas
($p=nT$) of $\pi^{-}$ with presumed conservation of particle
number to demonstrate that our hydro-kinetic approach agrees with
the main features of particle emission in the midrapidity region
of relativistic heavy ion collisions, known from results of many
transport code simulations. We also compare the particle momentum
spectra from HKM with ones obtained by means of the Cooper-Frye
prescription with standard isothermal freeze-out hypersurface.

We assume the following Bjorken-type initial conditions at
$\tau_{i}=1$ fm/c for HKM calculations: initial longitudinal
flow $v_{L}=z/t$ without transverse collective expansion,
ideal Boltzmann gas of $\pi^{-}$ in chemical (local)
equilibrium as the initial particle distribution,
boost-invariance of the system in the longitudinal direction
and cylindrical symmetry with Woods-Saxon initial energy
density profile in the transverse plane,
\begin{eqnarray}
  \epsilon(\tau_i,r_{T})={\epsilon_0\over\exp\left({r_T-R_T\over\delta}\right)+1}\,,
\label{num-1}
\end{eqnarray}
where $R_T=7.3$ fm, $\delta=0.67$ fm are associated with the
density profile of the \textit{Au} nucleus, and the maximal energy
density in the center of the system, $\epsilon_0=0.4$ GeV/fm$^3$,
corresponding to the temperature $T\approx 320 $ MeV of chemically
equilibrated pions. To find the momentum spectra in the HKM, the
relaxation time  $\tau_{\text{rel}}=p_{0} \tau^*_{\text{rel}} /
p^{\mu}u_{\mu}$ needs to be specified. For this aim we utilize the
expression given by Eq. (\ref{boltz-7}) that represents the rate
of binary collisions for one component Boltzmann gas. As for the
cross section, we carry out calculations for two distinct
(artificial) values: $\sigma=40$ mb and $\sigma=400$ mb. We
performed ideal hydro-calculations till $\tau_{f}=30$ fm/c when
the system becomes very rarefied, therefore the interactions are
nearly ceased and momentum spectra are almost frozen. The values
of $u_{\mu}(x)$, $T(x)$ and $\mu(x)$ near midrapidity,
$\eta\approx0$, are used then to calculate the emission function
$S( t^{\prime},\textbf{r}^{\prime},p)$, Eq. (\ref{emis-def}), and,
utilizing Eqs. (\ref{mom-sp2}) and (\ref{mom-sp3}) to evaluate the
momentum spectra at hypersurface $\tau = \tau_{f}\,$. Note that
the transformation of Eqs. (\ref{mom-sp2}) and (\ref{mom-sp3}) for
evolution parameter $\tau$ instead of $t$ with corresponding
substitution for the initial conditions can be easily done in a
straightforward manner.

\begin{figure}[h]
\centering
\includegraphics[scale=0.5]{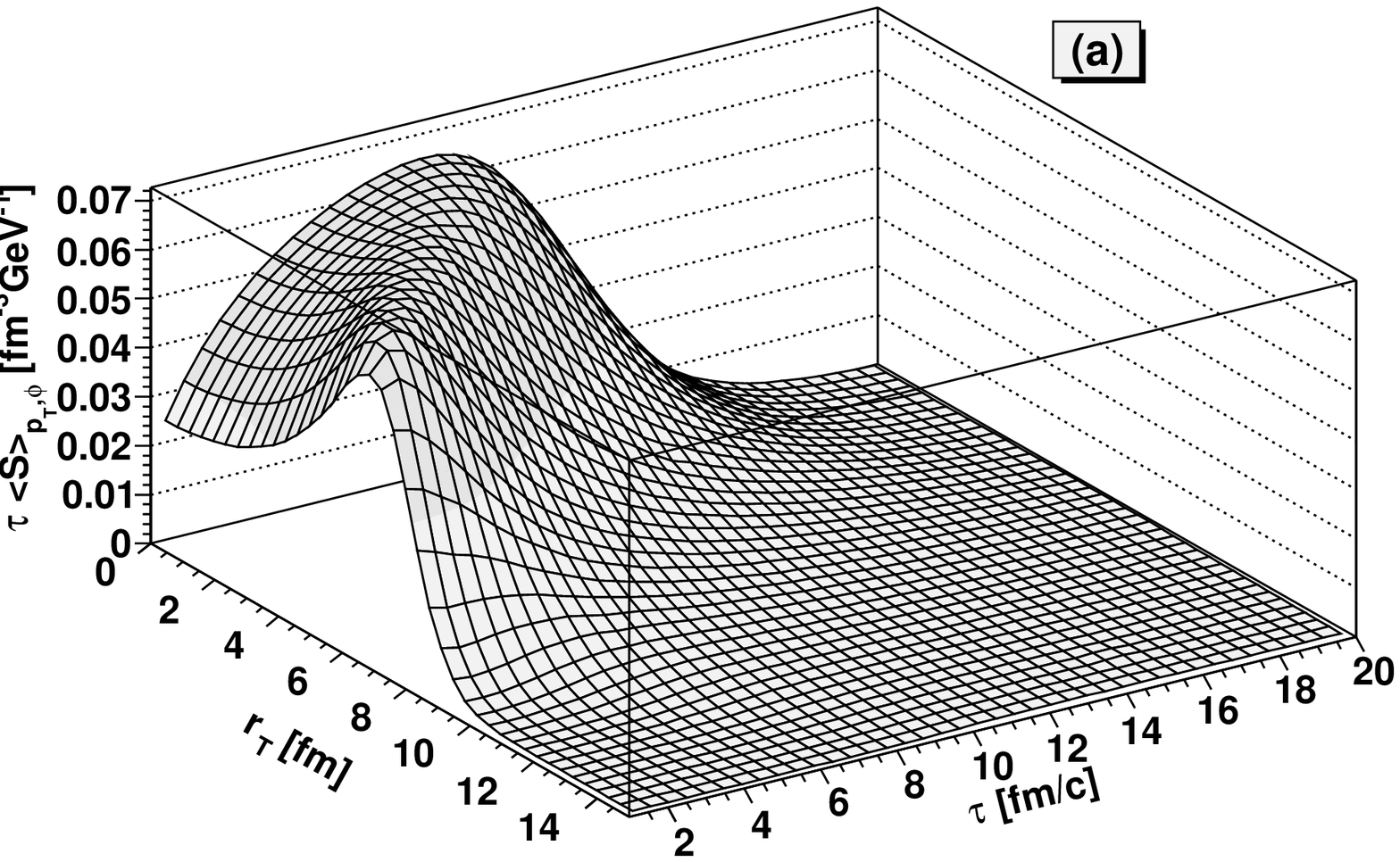}
 \includegraphics[scale=0.5]{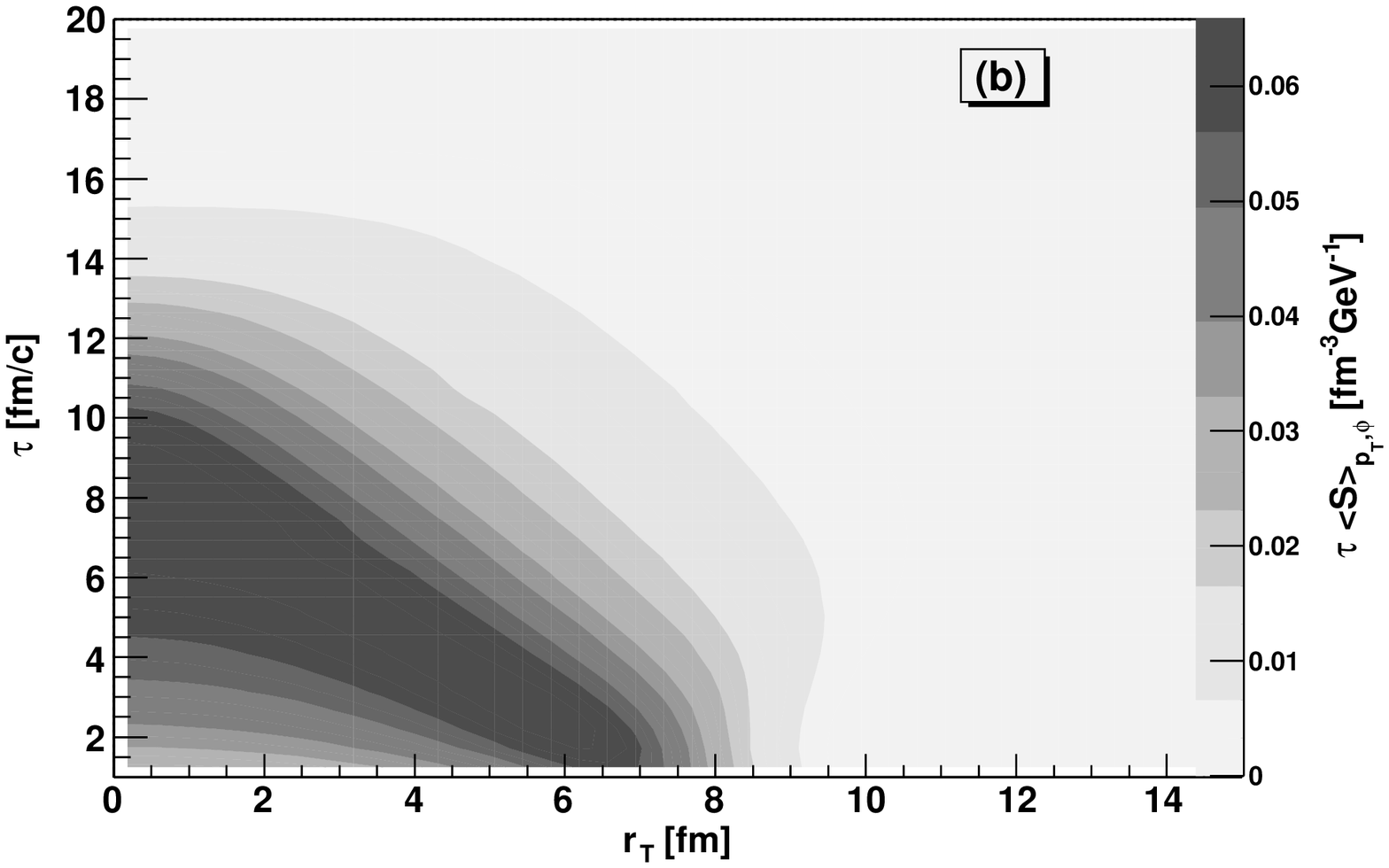}
\caption{Space-time dependence of the emission function integrated
over transverse momenta at $p_{L}=0$, for an expanding gas of
pions with cross section $40$ mb, initially with longitudinally
boost-invariant flow and Woods-Saxon energy density profile in the
transverse plane. } \label{40mb_proj}
\end{figure}

\begin{figure}[h]
\centering
\includegraphics[scale=0.5]{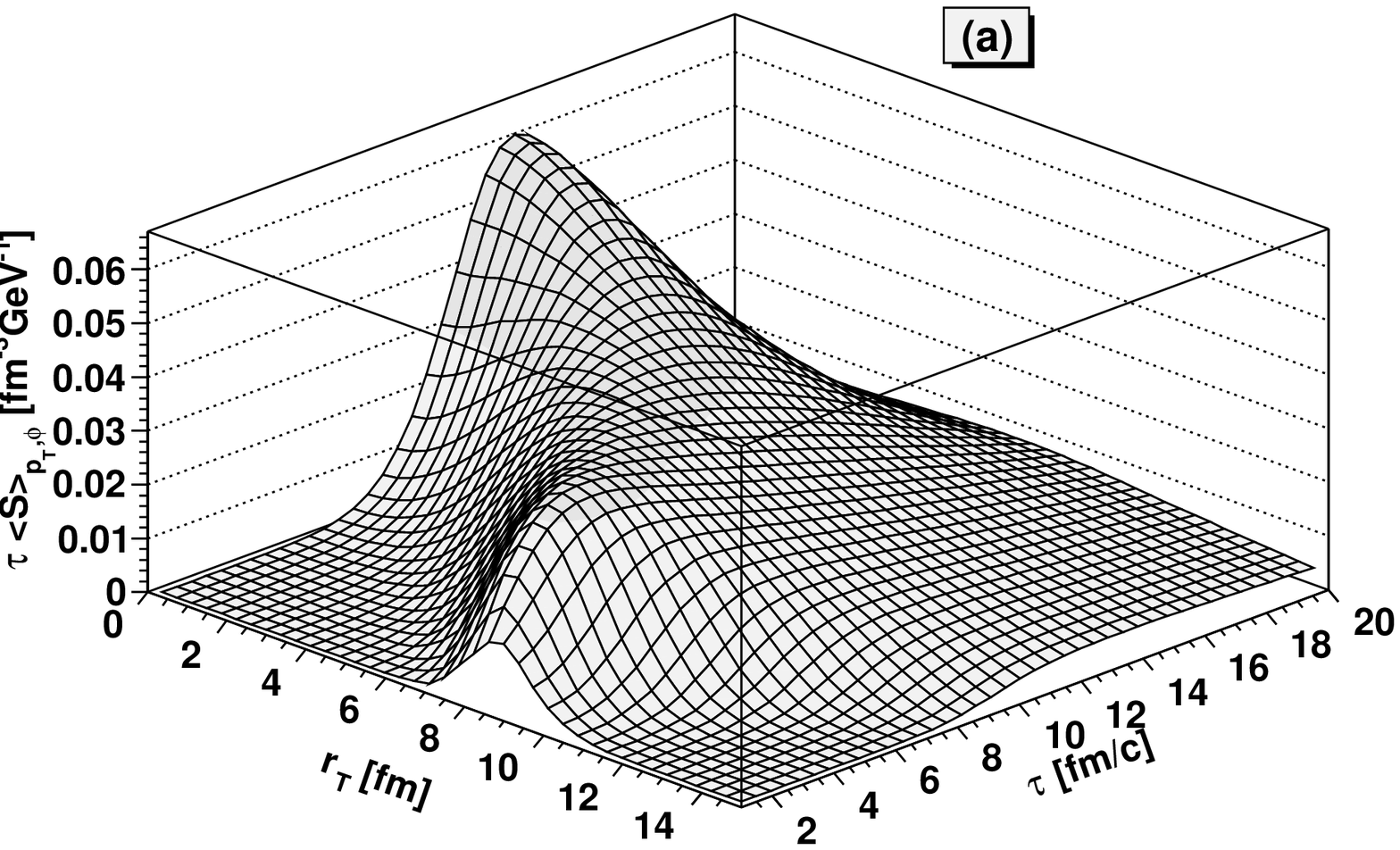}
\includegraphics[scale=0.5]{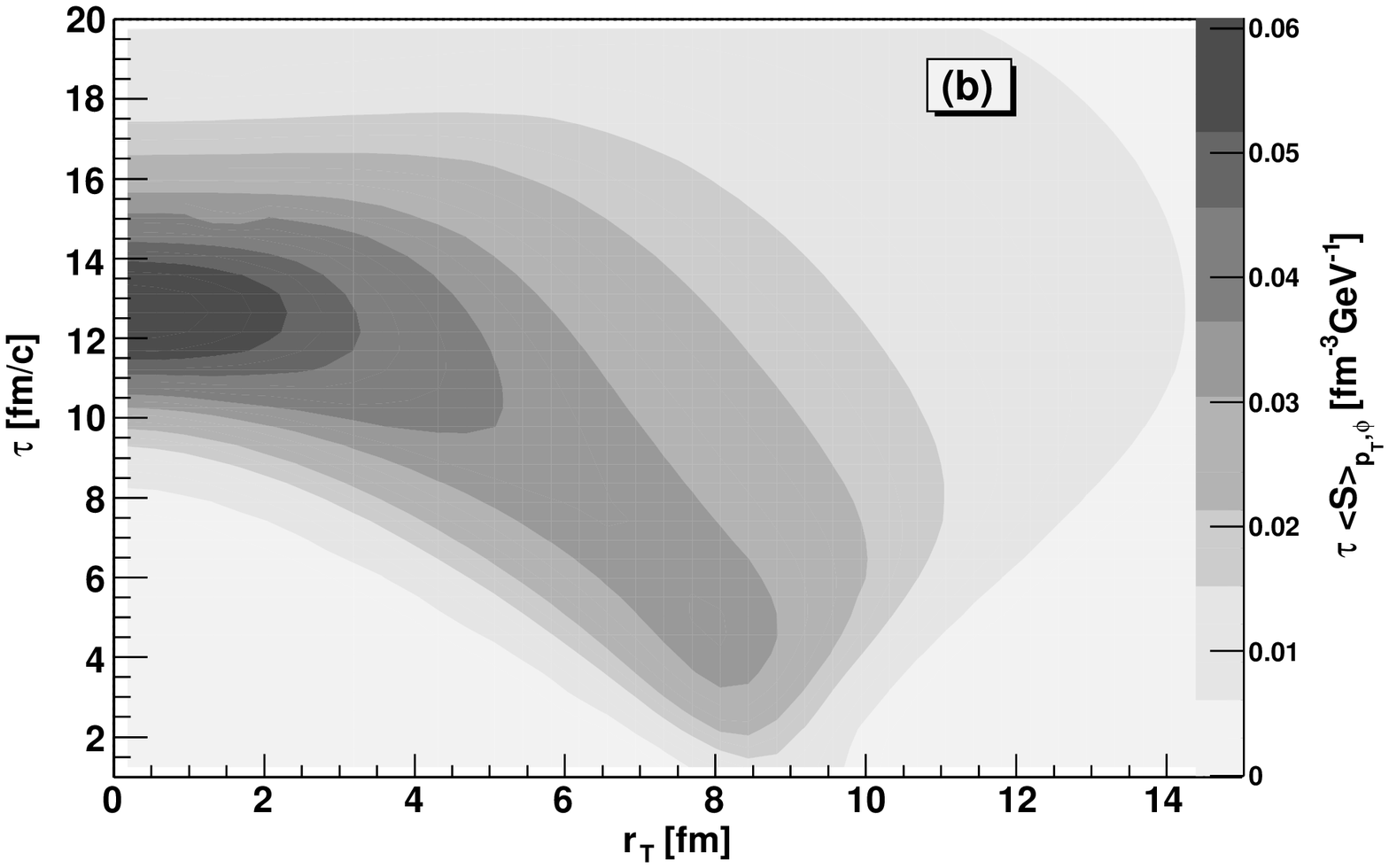}
\caption{ Same as in Fig. \ref{40mb_proj} but for a cross section
equal to $400$ mb.}
 \label{400mb_proj}
\end{figure}

\begin{figure}
    \centering
        \includegraphics[scale=0.5]{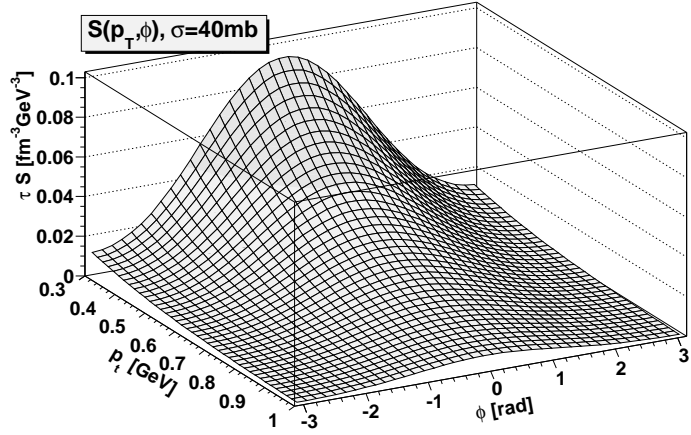}
        \caption{Emission function for an expanding gas of
        pions with cross section $40$ mb at $\tau=2$ fm/c
        as a function of the angle between the position and
        transverse momentum vectors of the escaping particle
        at the distance $r_T=6$ fm from the axis for
        different absolute values of transverse momentum.}
        \label{40mb_phi}
\end{figure}

\begin{figure}
    \centering
        \includegraphics[scale=0.5]{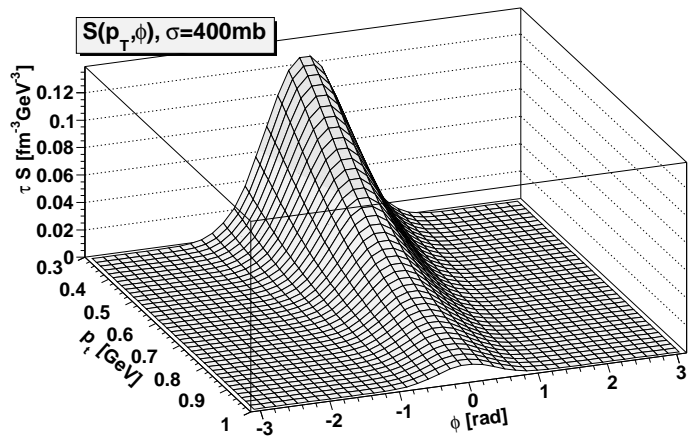}
        \caption{Same as in Fig. \ref{40mb_phi} but for
        cross section $400$ mb, $\tau=4$ fm/c and
        $r_T=8$ fm.}
        \label{400mb_phi}
\end{figure}

The results for the pion emission density integrated over
the transverse momenta
$\mathbf{p_T}=(p_{T}\cos\phi, p_{T}\sin \phi )$ at zero
longitudinal momentum, $p_{L}=0$,
\begin{eqnarray}
\langle S(x) \rangle_{p_{T},\phi}\equiv \int S(x,p)d^{2}p_{T},
 \label{num-0}
\end{eqnarray}
as a function of transverse radius $r_{T}$ and Bjorken proper time
$\tau$ are shown in Figs. \ref{40mb_proj} and \ref{400mb_proj} for
$\sigma=40$ mb and $\sigma=400$ mb,  respectively. The integrated
emission function is multiplied by a ``geometric" factor $\tau$,
because of kinematics in hyperbolic coordinates: $\tau\cdot
S(\tau, r_{T})$ is associated with the probability for particles
with any transverse momentum to be emitted in the central unit of
rapidity within the proper time interval ${[\tau,\tau+d\tau]}$ and
transverse radius $[r_T,r_T+dr_T]$. As one can see in these
figures,  the maximum of emission is more pronounced and occupies
narrower space-time area for larger cross sections than for
smaller ones, which is certainly a result of higher medium opacity
for higher rate of collisions. Therefore the freeze-out is rather
gradual process for the low ($\sigma=40$ mb) value of the cross
 section in the expanding gas, and one could hardly expect the
validity of the Cooper-Frye prescription there. Figures
\ref{40mb_phi} and \ref{400mb_phi} show the dependence of the
emission function $S$ on the angle between the position and
momentum vectors at the particle emission point in the transverse
plane.

\begin{figure}[h!]
    \centering
        \includegraphics[scale=0.5]{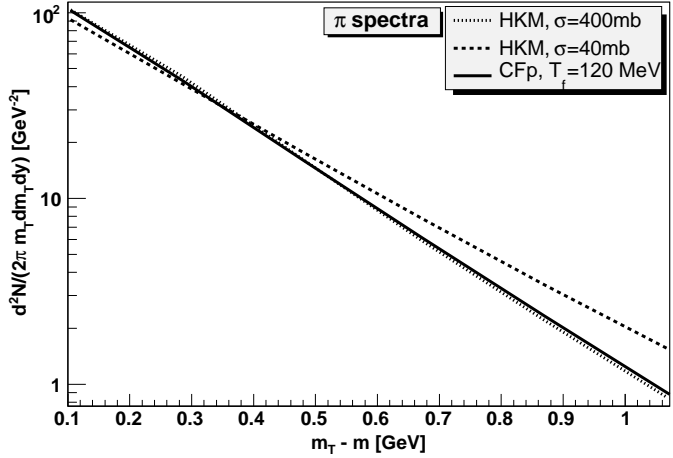}
    \caption{Transverse spectra of pions escaped until
    $\tau = 30$ fm/c from an expanding fireball calculated
    in HKM, with cross sections $40$ mb and $400$ mb, compared
   to the spectrum according to CFp applied to the local
   equilibrium  distribution at hypersurface $T_{f}=120$ MeV.}
    \label{spectra}
\end{figure}

To check the reliability of the conventional CFp applied at the
isotherm $T=T_{f}\sim m_{\pi}$ versus the hydro-kinetic picture of
the  continuous particle emission, we compared the transverse
momentum spectrum of emitted pions in HKM with the one calculated
in CFp. We performed an integration of the emission function over
a space-time four-volume till $\tau_f=30$ fm/c to obtain the
momentum spectrum of pions in the HKM. The emission of particles
already free at the initial moment $\tau_i$ was also accounted for
in such a calculation. On the other hand, we utilized CFp for an
isothermal hypersurface with $T_{f}=120 $ MeV and locally
equilibrated distribution function on it. The value of the
freeze-out temperature, $T_{f}=120 $ MeV,  is near temperatures
corresponding to maximum of the  emission function integrated over
transverse momenta  for an expanding gas of pions with
$\sigma=400$ mb. Such a hypersurface, as well as the velocity and
chemical potential distributions on it were extracted from a pure
hydrodynamic calculation. The results for $m_T$-spectra
($m_T=\sqrt{p_{T}^{2}+m^{2}}$) are presented in Fig.
\ref{spectra}. One can see the increase of the effective
temperature (inverse slope) in the case of small cross section
($\sigma=40$ mb)  compared to pure hydrodynamic calculations with
Cooper-Frye prescription. This happens because, if the collision
rate is small, particles can escape easily from the early stages
of the evolution when $T>T_{f}$ (see Fig. \ref{40mb_phi}). While
the transverse collective velocity for particles that escape at
higher temperatures $T>T_{f}$ is smaller than the one for
particles which suffer freeze out according to CFp at $T=T_{f}$,
the resulted gain in collective velocity does not lead, typically,
to an increase of the effective temperature calculated according
to ideal hydro-equations because of the energy transfer from
transverse to longitudinal degrees of freedom.  Then the
utilization of CFp with $T_{f}\sim m_{\pi}$ leads to a noticeable
decrease of the inverse slope compared to the hydro-kinetic result
(see Fig. \ref{spectra}). Apparently, the high collision rate that
occurs for $\sigma=400$ mb prevents an early escaping and,
consequently, leads to a gradual freeze-out in the rather dilute
medium at low temperatures $T<T_{f}$. The interactions at this
stage do not change the momentum spectrum essentially justifying,
thereby, the utilization of CFp for the $T_{f}=120 $ MeV isotherm.

Now, let us consider a more sophisticated hydro-kinetic model
accounting for some realistic features of the evolution of
fireballs created in ultrarelativistic heavy ion collisions. We
focus on the midrapidity hadronic emission at RHIC energies. Note
that resonance decay contributions to pion spectra are not taken
into account in the present version of HKM. Therefore, we do not
compare our results with experimental data. Our aim here is to
study the influence of different types of initial conditions and
equations of state on hadronic emission processes. Special
attention is paid to the lifetime of the system that undergoes a
phase transition, because, as is well known, the long evolution
time is one of the main shortcomings of hydrodynamic and kinetic
models aiming to describe the quark-gluon plasma (QGP) to hadron
gas (HG) transition in relativistic heavy ion collisions: it leads
to an overestimate of the the corresponding lifetime scale found
in HBT analysis \cite{puzzle}.

 First of all, to solve the relativistic hydrodynamic
equations, an equation of state needs to be specified. Because the
thermal model analysis of the particle number ratios at RHIC
demonstrates almost zero baryon chemical potential \cite{PBM}, we
utilize an  EoS with zero net baryon density. According to lattice
QCD calculations, the deconfinement transition at a vanishing
baryon chemical potential is a rapid crossover rather than a first
order phase transition with singularities in the bulk
thermodynamic observables \cite{Karsch}. Therefore for the
calculations presented here, we use for high temperatures a
realistic EoS \cite{Laine} adjusted to the QCD lattice data with a
crossover transition at about $T_{\text{c}}\approx175$ MeV and
matched with an ideal chemically equilibrated multicomponent
hadron resonance gas at $T=T_{\text{c}}$. For temperatures in the
interval $T_{\text{ch}}<T<T_{\text{c}}$, $T_{\text{ch}}=160$ MeV,
we utilize an EoS \cite{Laine} of an ideal chemically equilibrated
multicomponent hadron resonance gas, and for $T<T_{\text{ch}}$ we
obtain and utilize an EoS of a multicomponent \cite{pdg} hadron
resonance gas with a chemical composition frozen at
$T=T_{\text{ch}}$. The resonance mass spectrum extends over all
mesons and baryons with masses below $2.6$ GeV \cite{pdg};
corresponding electronic tables where taken from the FASTMC event
generator \cite{Amelin}. The value of the chemical freeze-out
temperature, $T_{\text{ch}}=160$ MeV, is chosen because it is near
chemical decoupling temperatures \cite{PBM} extracted from the
measured hadron abundance ratios at RHIC.\footnote{Note to avoid
the  misunderstanding that the decay of resonances during the
hydrodynamic expansion of hadronic matter agrees with the chemical
freeze-out concept but could influence an EoS and particle
spectra, we neglect here this effect just for simplicity reasons.}

Then, to calculate the equations of state
$p=p(T,\mu_{1},...,\mu_{n})$, $\epsilon=\epsilon
(T,\mu_{1},...,\mu_{n})$ of the chemically frozen ideal hadron
resonance gas, one needs to know how the chemical potential
$\mu_{i}$ of each particle species and energy density depend on
temperature when the hadron gas, being in an initially {\it known}
state (all $T_{\text{ch}}, \epsilon_{\text{ch}},\mu^i_{\text{ch}}$
are known), expands adiabatically. It can be done in the following
way. Firstly, note that the concentration of any particle species
$``i"$,
\begin{eqnarray}
 \kappa_{i}=\frac{n_{i}(T,\mu_{i})}{n(T,\mu_{1},...,\mu_{n})}, \quad n=\sum n_{i},  \label{num-en01}
 \end{eqnarray}
 is fixed by its initial value at $T_{\text{ch}}$ and, so,
\begin{eqnarray}
\frac{n_{i}(T,\mu_{i})}{n_{j}(T,\mu_{j})}=\frac{\kappa_{i}}{
\kappa_{j}} \label{num-en02}
 \end{eqnarray}
does not change with temperature for $T<T_{\text{ch}}$. Here
$n_{i}$ is the number density of particle species ``$i$" and, in
Boltzmann approximation,
\begin{eqnarray}
 n_{i}=\frac{(2j_{i}+1)}{2\pi ^{2}}T m_{i}^{2}\exp
(\mu _{i}/T)K_{2}(m_{i}/T),
\label{num-en03}
\end{eqnarray}
 where $K_{n}(u)=\frac{1}{2}\int_{-\infty }^{+\infty }dz\exp [-u\cosh
z+nz]$ , with $\mathop{\rm Re} u>0,$ is the modified Bessel
function of order $n$ ($n=0,1,...$). The  energy density and
pressure of the chemically frozen mixture of ideal Boltzmann gases
are
\begin{eqnarray}
  \epsilon = n (T,\mu_{1},...,\mu_{n}) \sum \kappa_{i} e_{i}(T), \quad p(T,\mu_{1},...,\mu_{n})T,
  \label{num-en1}
\end{eqnarray}
where
\begin{eqnarray}
  e_{i}\equiv
  \frac{\epsilon_{i}}{n_{i}}=m_{i}\frac{K_{1}(m_{i}/T)}{K_{2}(m_{i}/T)}+
  3T.
\label{num-en2}
\end{eqnarray}
 Then, taking into account the thermodynamic identity
\begin{eqnarray}
\epsilon + p = sT + \sum \mu_{i}n_{i} \label{num-en3}
\end{eqnarray}
and, accounting for the constancy of entropy density to particle
number density ratio, $s/n$, during an ideal hydro-evolution, we
divide the above identity by $(nT)$ and get
\begin{eqnarray}
\sum \kappa_{i} \left (\frac{e_{i}(T)}{T}  -
\frac{e_{i}(T_{\text{ch}})}{T_{\text{ch}}}\right )= \sum
\kappa_{i} \left (\frac{\mu_{i}(T)}{T}  -
\frac{\mu_{i}(T_{\text{ch}})}{T_{\text{ch}}}\right ).
\label{num-en4}
\end{eqnarray}
Equations (\ref{num-en4}) and (\ref{num-en02}), (\ref{num-en03})
define $\mu_{i}(T)$ for all particle species ``$i$" and, thereby,
complete the thermodynamic trajectory (for $T<T_{\text{ch}}$) of
the system and all their components: $\epsilon(T), p(T), \mu_i(T)$
in isentropic expansion starting from $T_{\text{ch}}$. Of course,
the equations of state of the chemically frozen hadron resonance
gas (\ref{num-en1}) can be reduced now to $p(\epsilon)$ on this
{\it particular}  thermodynamic trajectory.

\begin{figure}[h!]
    \centering
        \includegraphics[scale=1.0]{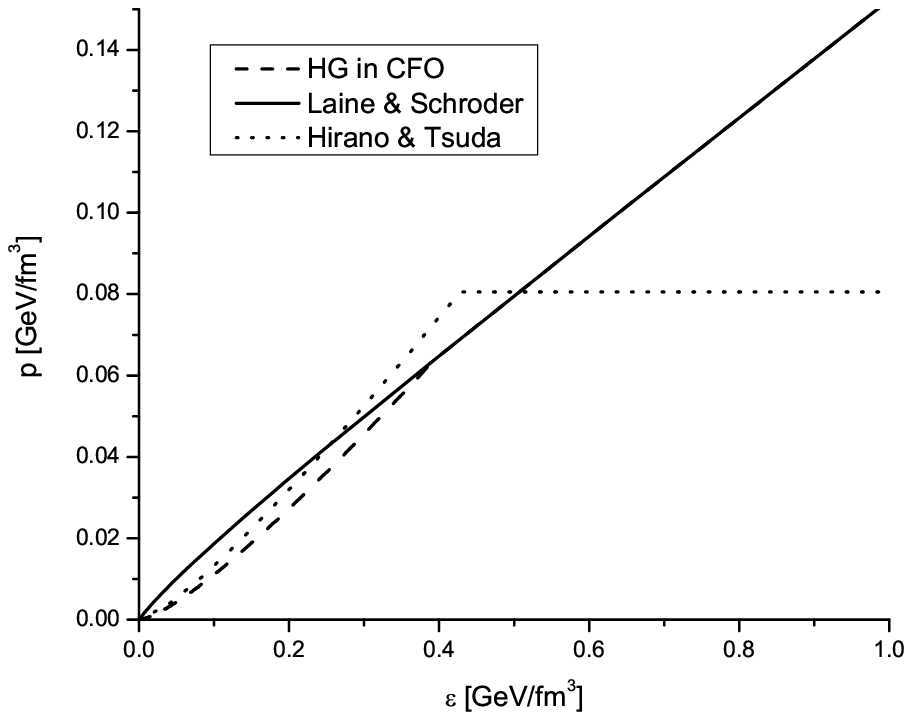}
    \caption{Pressure versus energy density for Laine and
    Schr\"{o}der EoS with crossover transition; for
    Hirano and Tsuda EoS with strong first order phase
    transition; and for chemically frozen ideal hadron
    resonance gas.}
    \label{fig_e-p}
\end{figure}

\begin{figure}[h!]
    \centering
        \includegraphics[scale=1.0]{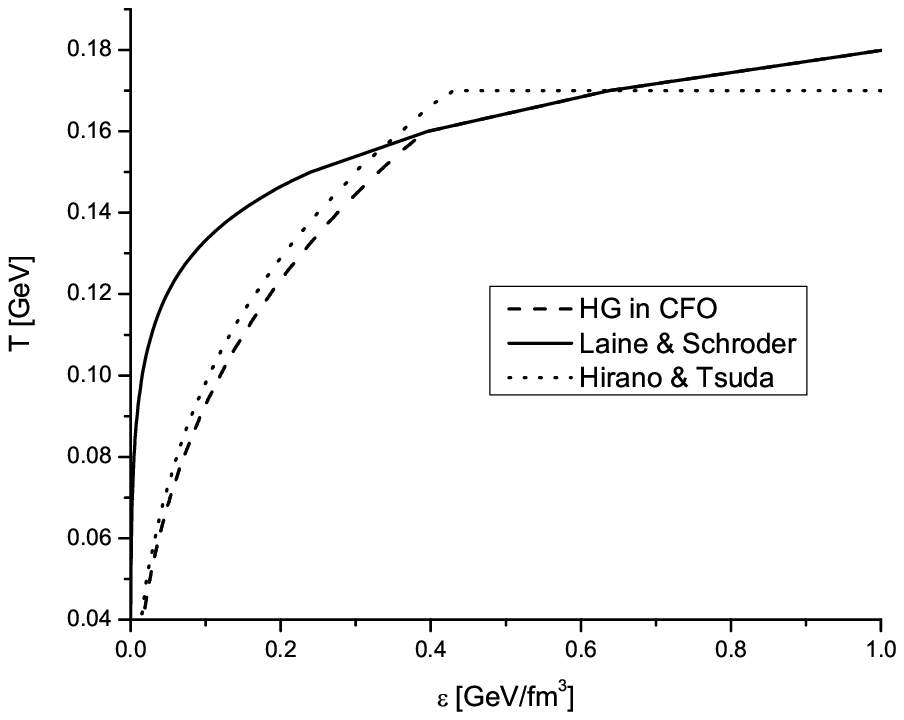}
    \caption{Temperature versus energy density for Laine and
    Schr\"{o}der EoS with crossover transition; for Hirano
    and Tsuda EoS with strong first order phase transition;
    and for chemically frozen ideal hadron resonance gas.}
    \label{fig_e-t}
\end{figure}

\begin{figure}[h!]
    \centering
        \includegraphics[scale=1.0]{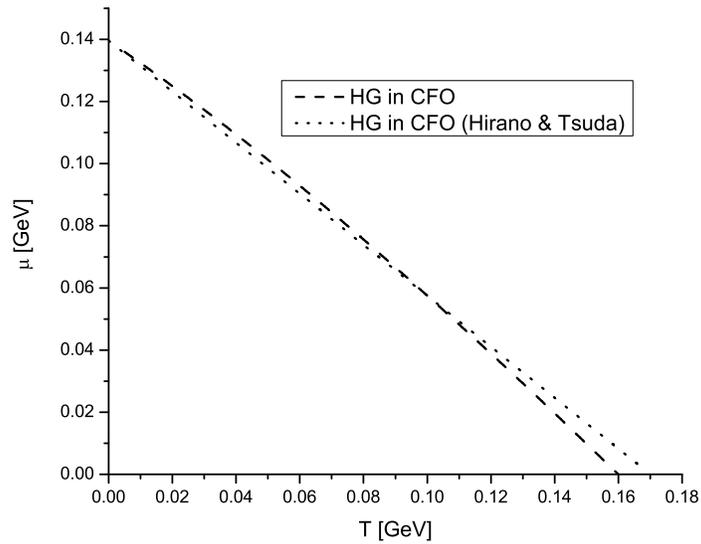}
    \caption{Chemical potential of pions (in the Boltzmann
    approximation) versus energy density for Laine and
    Schr\"{o}der EoS with crossover transition; for Hirano
    and Tsuda EoS with strong first order phase transition;
    and for chemically frozen ideal hadron resonance gas.}
    \label{fig_e-mu}
\end{figure}

For comparison, we also perform ideal hydro-calculations with the
EoS taken from Ref. \cite{Hirano1}, where a strong first order
phase transition is assumed. The corresponding EoS are presented
in Figs. \ref{fig_e-p}, \ref{fig_e-t},  \ref{fig_e-mu}.

To calculate the pionic emission function and momentum spectra by
means of HKM, the rate of collisions needs to be specified. We use
the expression for the collision rate (in the co-moving system)
that accounts for binary collisions of  pions with any ``$i$"
hadronic species,
\begin{eqnarray}
\frac{1}{\tau_{\text{rel}}^{\text{id}*}(x,p)} = \sum  \int d^3k
\frac{g_{i}}{(2\pi)^3} \exp\left(-\frac{E_{k,i} -
\mu_{i}(x)}{T(x)}\right )  \sigma_{i}(s)
\frac{\sqrt{(s-(m-m_{i})^2)(s-(m+m_{i})^2)}}{2 E_{p}E_{k,i}}.
\label{num-en5}
\end{eqnarray}
Here $g_{i}=(2j_{i}+1)$, $E_{p}=\sqrt{\mathbf{p}^2+m^2}$,
$E_{k,i}=\sqrt{\mathbf{k}^2+m_{i}^2}$,  $s=(p+k)^{2}$ is the
squared c.m. energy of the pair, and $\sigma_{i}(s)$ is the total
cross section in the corresponding binary collision. For the
latter, we utilize a Breit-Wigner resonance formula with
$\sqrt{s}$-dependent parametrization of decay widths as in Ref.
\cite{Bass}. All relevant resonance states from \cite{pdg}, used
in the FASTMC event generator \cite{Amelin} -  $359$ different
species - are taken into account for the calculation of
$\sigma_{i}(s)$. Suppression of bulk pionic emission from very
high energy density stage of matter evolution where quark-gluon
degrees of freedom are dominated is assured in our approach by
abrupt increase of the collision rate (\ref{num-en5}) that is
conditioned by a drastic increase of hadronic density at $T
\gtrsim T_{\text{c}}$ in Eq. (\ref{num-en5}). Note, to avoid a
misunderstanding, we utilize the relaxation rate (\ref{num-en5})
to calculate the pionic emission for $T>T_{\text{c}}$ for the sake
of simplicity only. In a more advanced approach the gradual
disappearance of pions and other hadronic degrees of freedom
during the crossover transition as well as the scattering of pions
with quarks, gluons, etc., should be taken into account. Here we
just expect that it will lead to the same effect of drastic
suppression of the pionic emission from the region occupied by
QGP.

As for the initial conditions, we first perform calculations with
a Woods-Saxon initial energy density profile (\ref{num-1}) in the
transverse directions at $\tau_i=1$ fm/c. The values of parameters
we choose for illustrative calculations are $\epsilon_0=6$
GeV/fm$^3$, $R_T=7.3$ fm, and $\delta=0.67$ fm. It results in the
initial temperature at the central ``plateau" $T=247$ MeV. These
initial conditions at midrapidity are similar to the ones used in
Ref. \cite{Hirano-Morita}, where the transverse energy density
distribution at $\tau_{i}=1$ fm/c was parametrized by a flat
region with Gaussian smearing near the edge. The hydro-evolution
starts at $\tau_i=1$ fm/c with initial Bjorken flow without
transverse velocity ($v_T=0$), assuming a complete chemical
equilibrium at $T>T_{\text{ch}}$ and a chemically frozen evolution
below $T_{\text{ch}}$.

\begin{figure}[h!]
    \centering
        \includegraphics[scale=0.5]{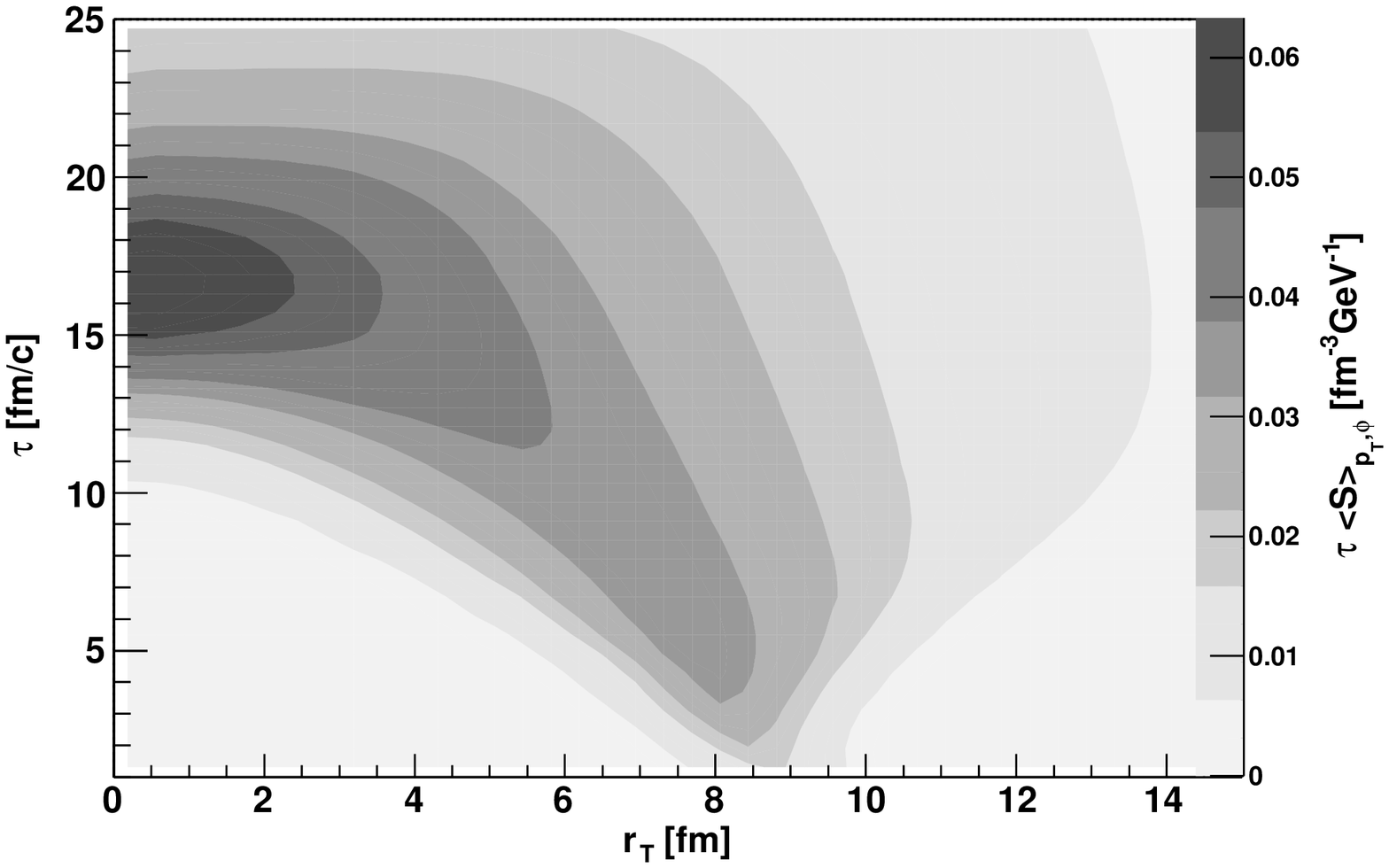}
    \caption{Space-time dependence of the pionic emission
    function integrated over the transverse momenta for an
    expanding fireball with  crossover transition,
    initially with longitudinally boost-invariant flow and
    Woods-Saxon energy density profile in the transverse
    plane.}
    \label{fig_S_real}
\end{figure}

\begin{figure}[h!]
    \centering
        \includegraphics[scale=0.5]{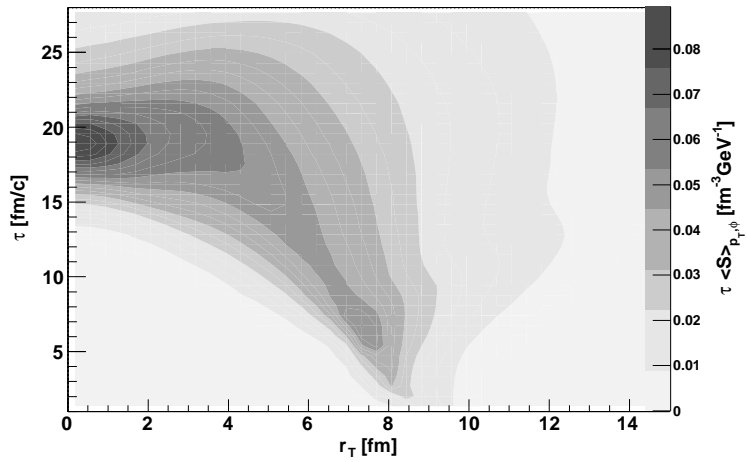}
    \caption{Same as in Fig. \ref{fig_S_real} but for a strong
     first order phase transition.}
    \label{fig-H}
\end{figure}

\begin{figure}[h!]
    \centering
        \includegraphics[scale=0.45]{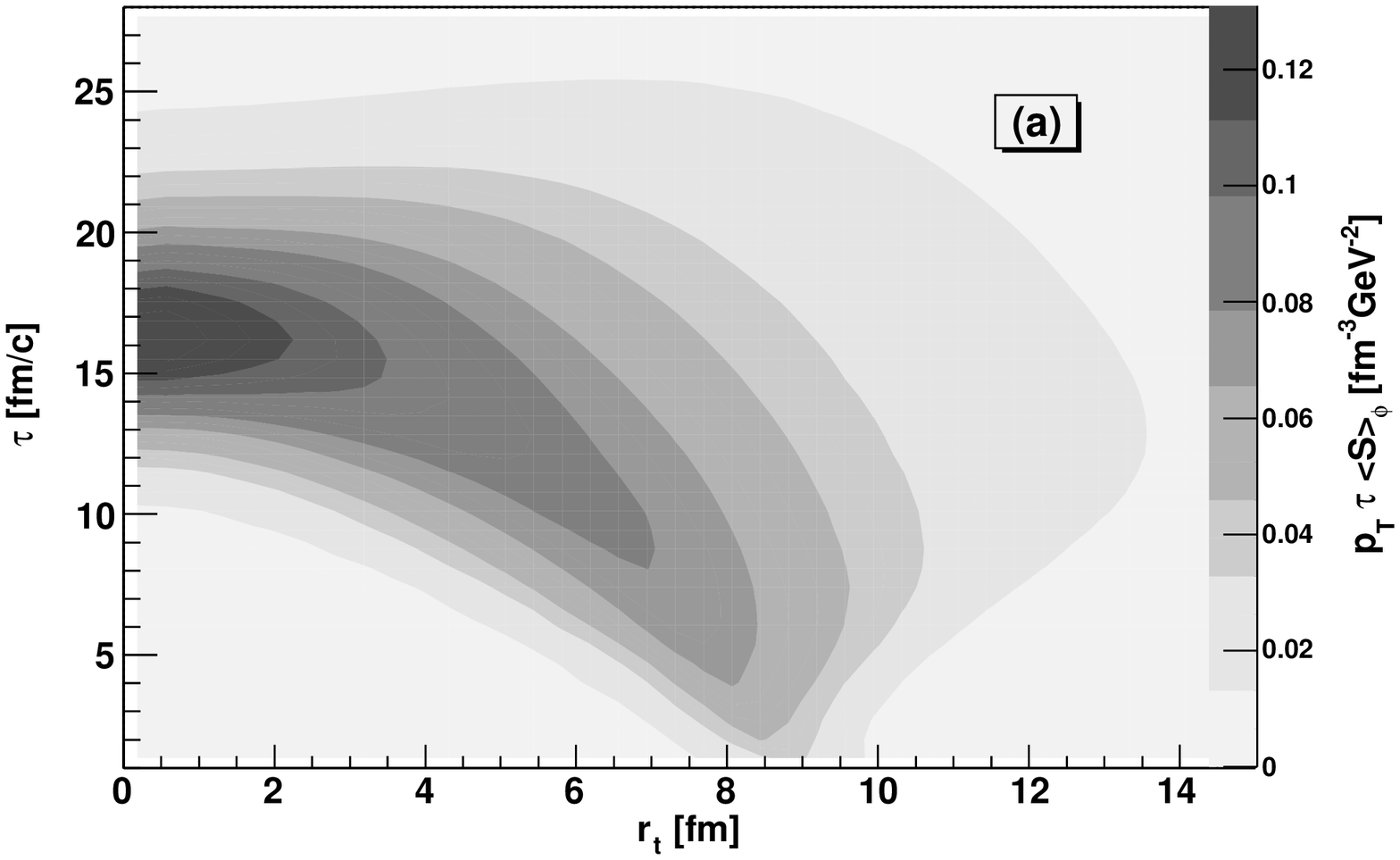}
         \includegraphics[scale=0.45]{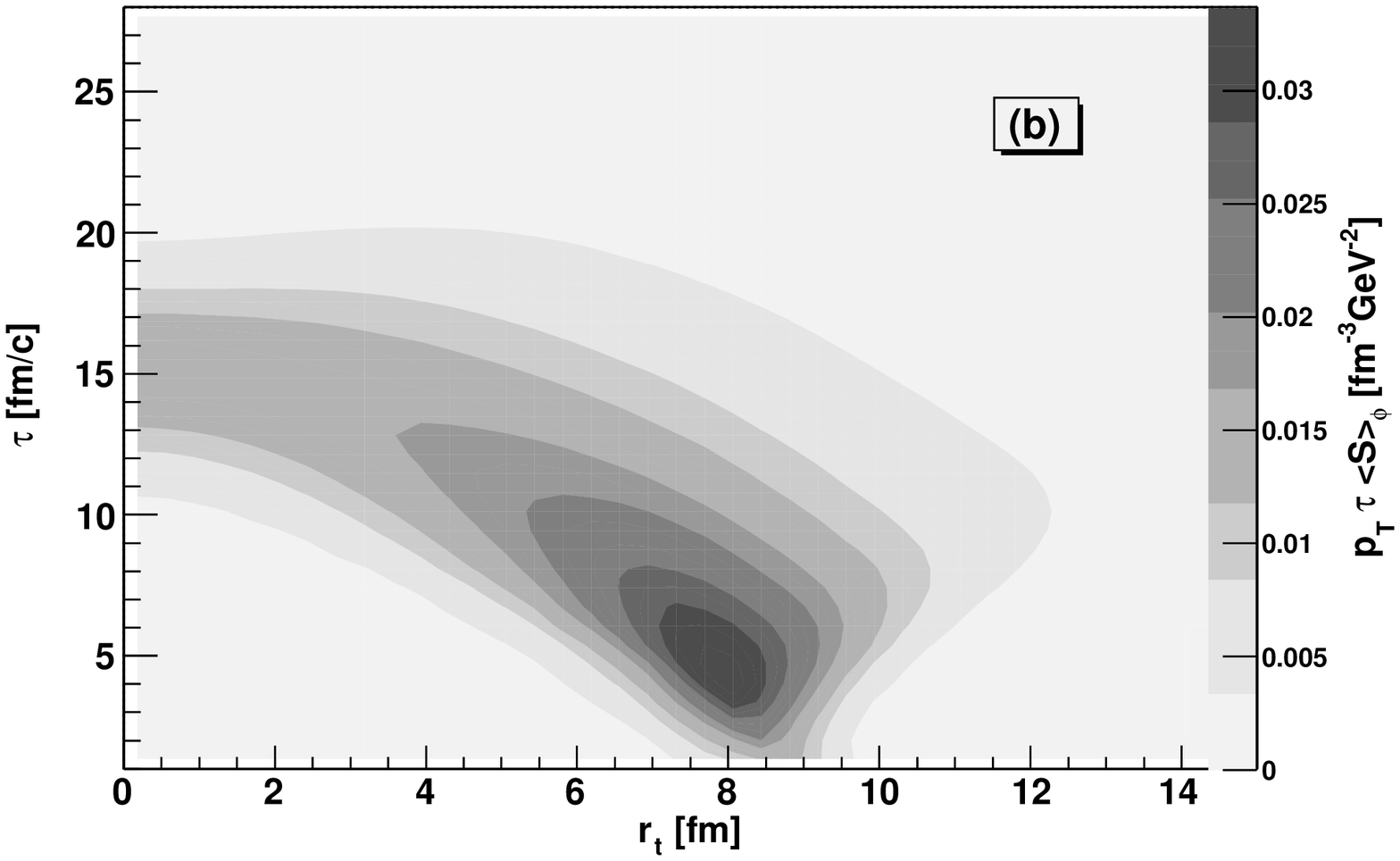}
          \includegraphics[scale=0.45]{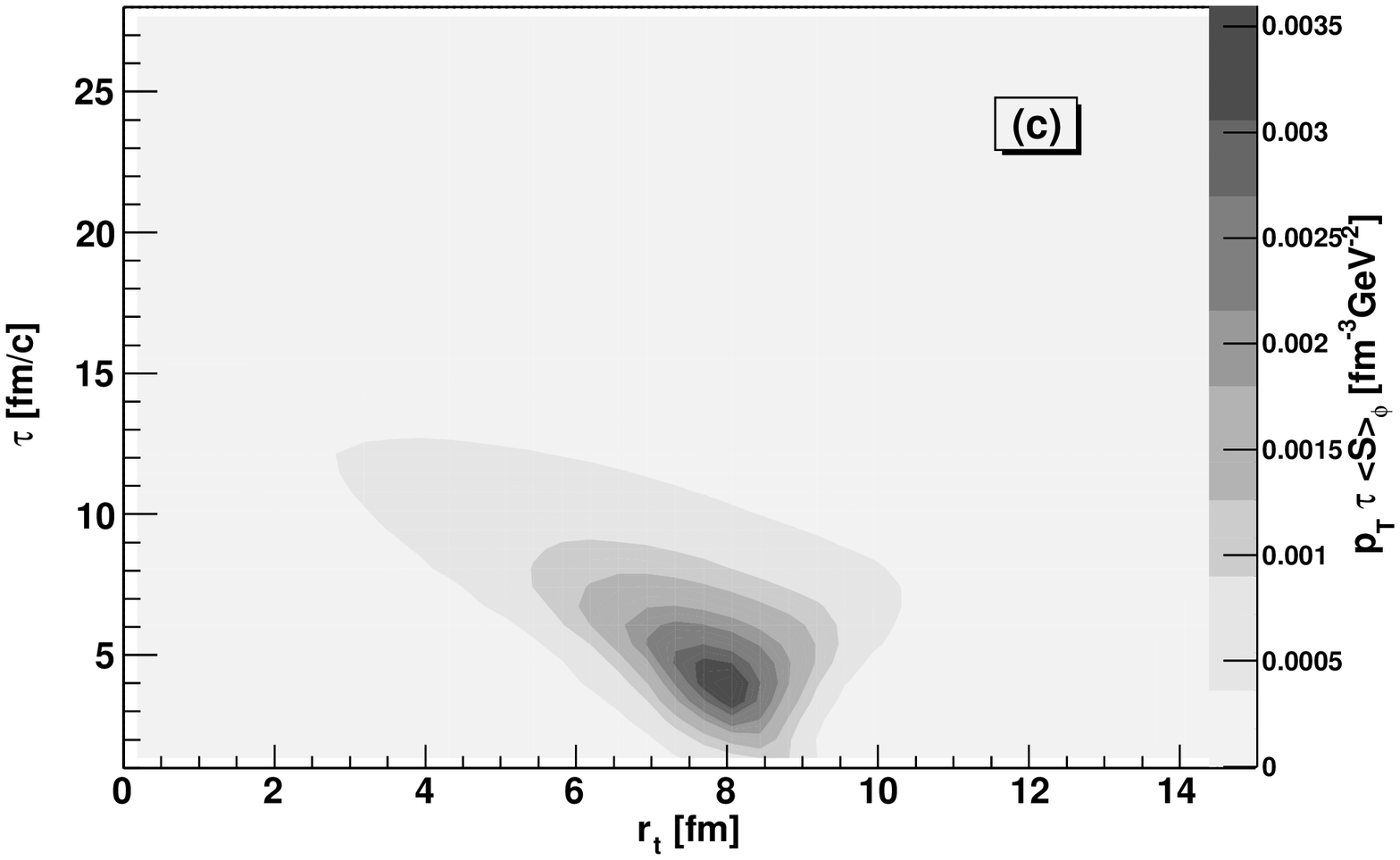}
    \caption{Space-time dependence of the pionic emission
    function of an expanding fireball with crossover
    transition and Woods-Saxon initial energy density
        profile in the transverse plane, at $\phi=0$ and
        different values of $p_{T}$: (a) $p_{T}=0.3$ GeV, (b) $p_{T}=0.6$ GeV, (c) $p_{T}=1.2$ GeV.}
   \label{diff-pt1}
\end{figure}

The $\mathbf{p_T}$-integrated pion emission function at
midrapidity is shown in Fig. \ref{fig_S_real} for an EoS with a
crossover transition and in Fig. \ref{fig-H} for an EoS with
strong first order phase transition taken from Ref.
\cite{Hirano1}. One can see that the utilization of the EoS with a
crossover transition leads to a significant decrease of the system
lifetime because of a more effective acceleration of the system
\cite{Hama3} in expansion and, therefore, is more reliable in view
of RHIC data indicating a rather short lifetime of the system
\cite{puzzle}.\footnote{The same conclusion was reached recently
in Ref. \cite{Flor} based on the analysis of effects of different
forms of the sound velocity function on the hydrodynamic evolution
of matter created in ultrarelativistic heavy ion collisions.}
Results for a non-$\mathbf{p_T}$-integrated emission function are
presented in Fig. \ref{diff-pt1}. They demonstrate the tendency of
high-$p_T$ particles to be emitted early from the periphery of the
system, whereas low-$p_T$ particles are mostly emitted at the late
stage of evolution from the center of the system when the system
becomes fairly rarefied because of the expansion. It corresponds
to results of Ref. \cite{Molnar}, where the same conclusion was
drawn based on partonic cascade model calculations.

\begin{figure}[h!]
    \centering
        \includegraphics[scale=0.5]{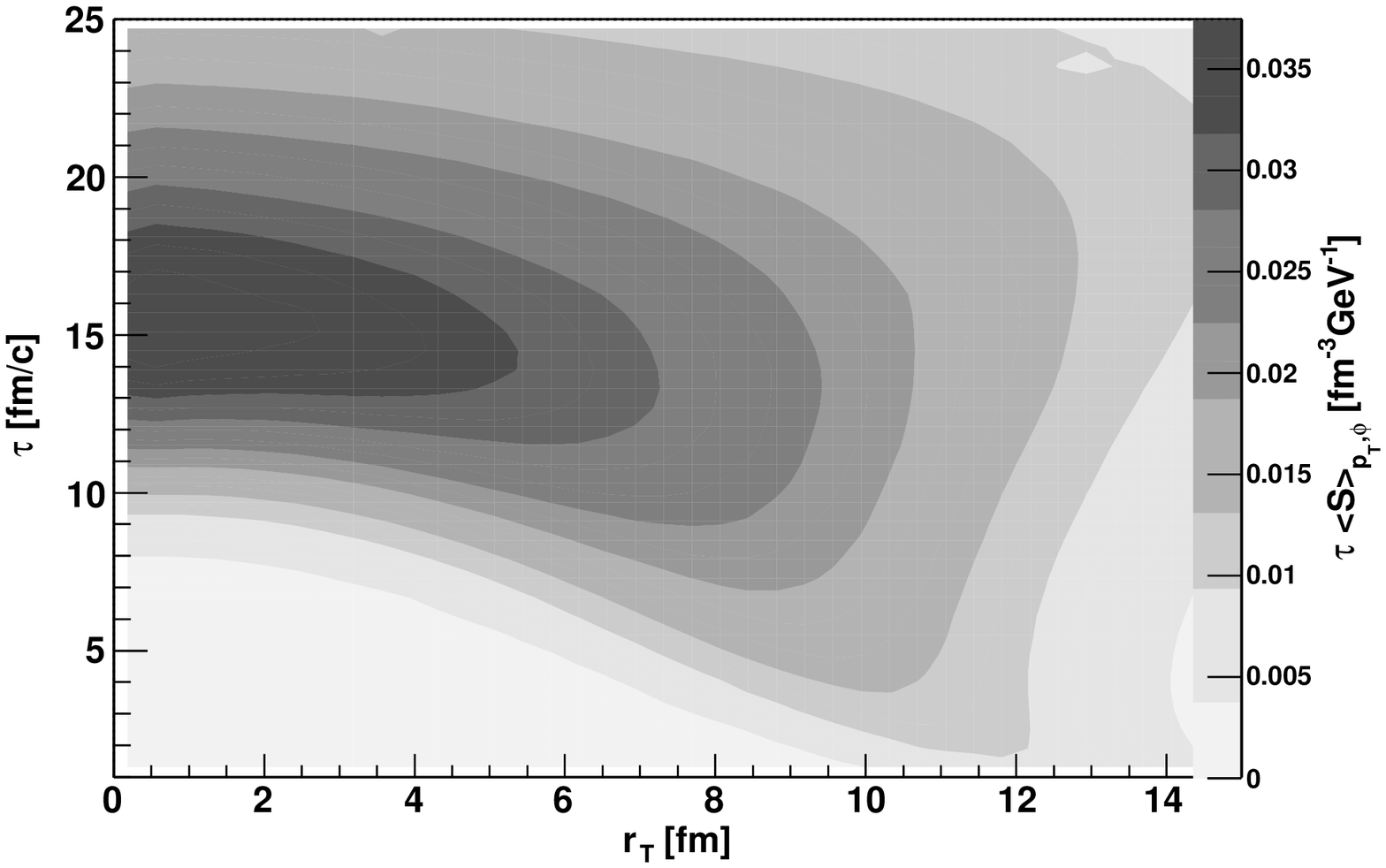}
    \caption{Space-time dependence of the pionic emission
    function integrated over transverse momenta for an
    expanding fireball with crossover transition, initially
    with longitudinally boost-invariant flow and Gaussian
    energy density profile in the transverse plane.}
    \label{fig_s_real}
\end{figure}

\begin{figure}[h!]
    \centering
        \includegraphics[scale=0.5]{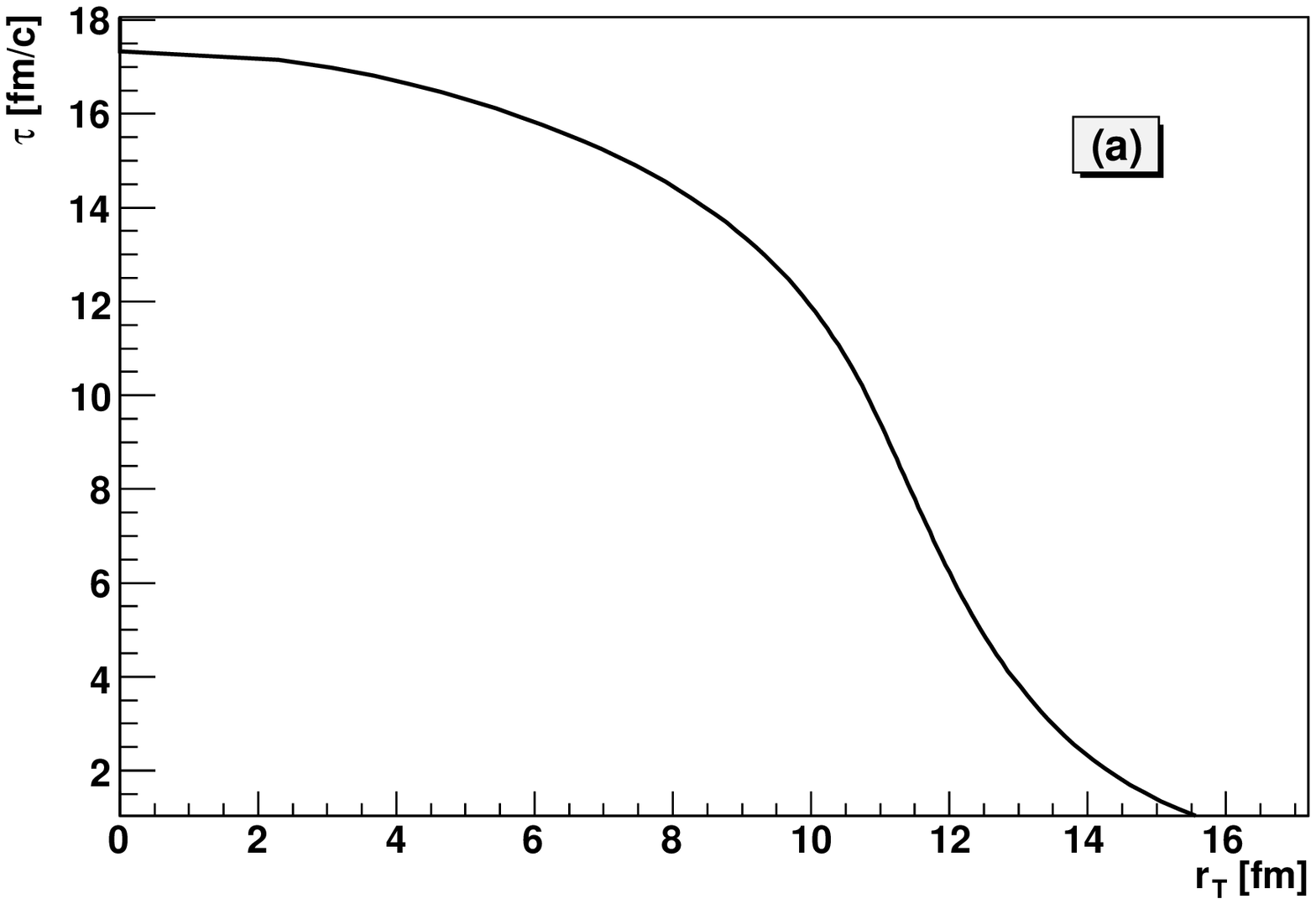}
         \includegraphics[scale=0.5]{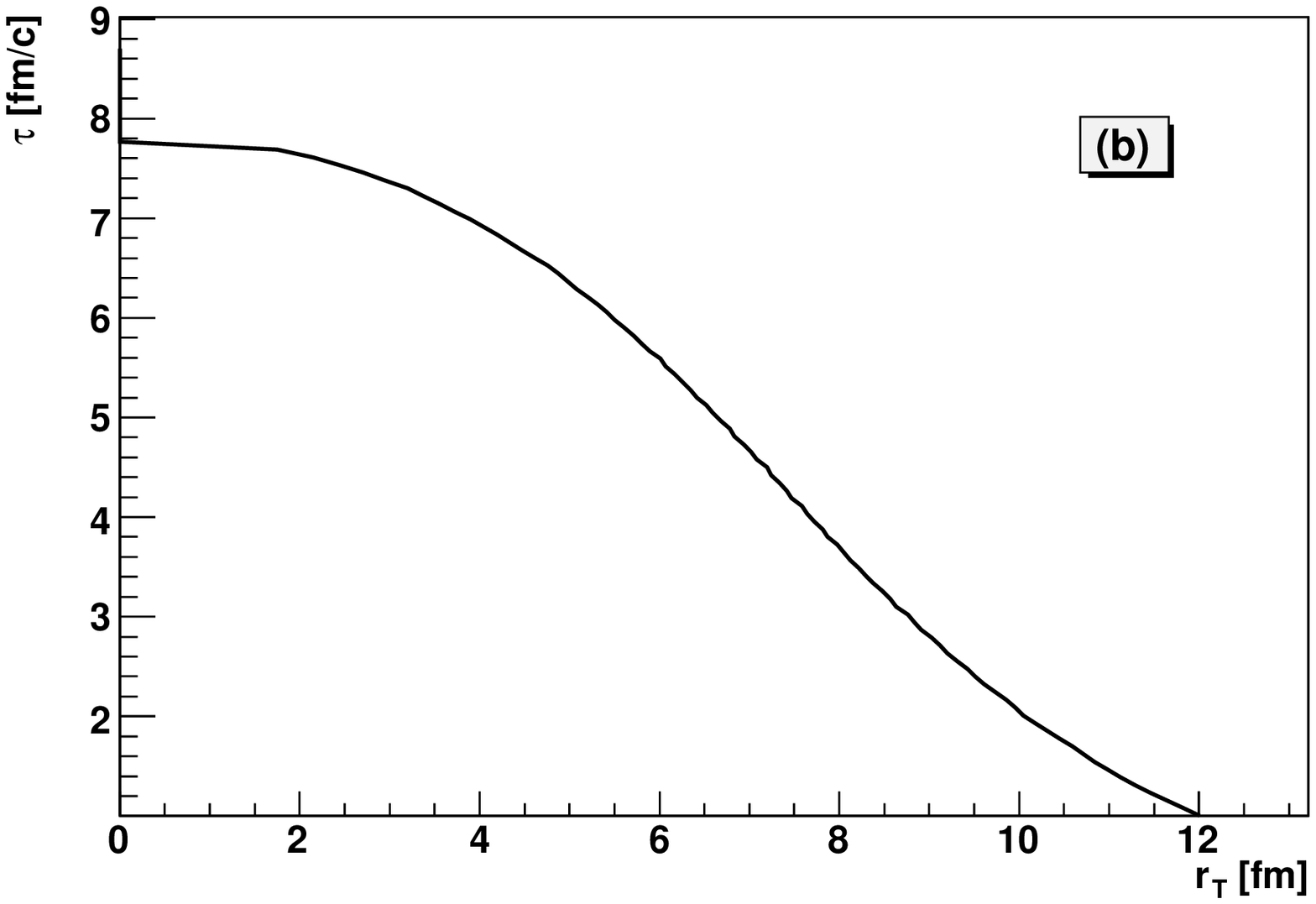}
    \caption{Isothermal hypersurfases of an expanding
    fireball with crossover transition and Gaussian
    initial energy density profile in the transverse plane: (a) $T_{f}=75$ MeV, (b) $T_{f}=160$ MeV. }
    \label{fig_freezeout_real}
\end{figure}

\begin{figure}[h!]
    \centering
        \includegraphics[scale=0.5]{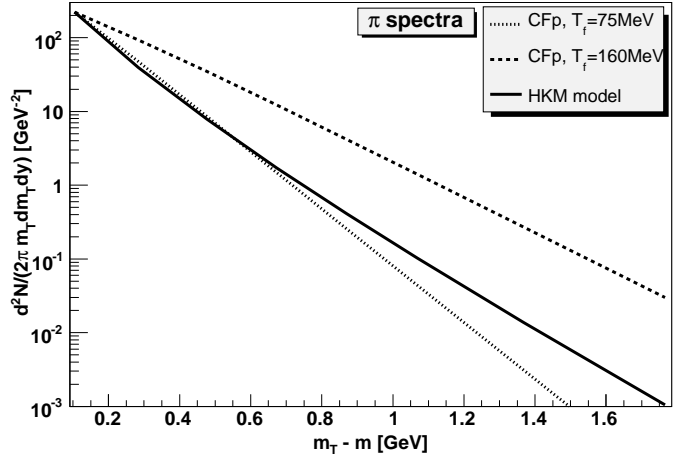}
    \caption{Transverse spectrum of pions escaped until
    $\tau = 30$ fm/c from an expanding fireball with
    crossover transition calculated in HKM versus spectra
    calculated according to CFp applied to the local equilibrium
    distribution at $T_{f}= 75$ MeV and $T_{f}=160$ MeV.}
    \label{fig_spectra_real}
\end{figure}

While the system lifetime is reduced if the EoS with a crossover
transition is utilized, it could still be too high for reproducing
measured values of the HBT radii at RHIC \cite{puzzle}. To study
the influence of the initial conditions on the lifetime of the
system, we perform hydro-kinetic calculations also for a Gaussian
initial energy density profile, instead of a Woods-Saxon one:
\begin{equation}
\epsilon(\tau_i, r_T)=\epsilon_0\exp\left(-{r_T^2\over
R_T^2}\right), \label{num-2}
\end{equation}
where $R_T=7.3$ fm and the normalization constant is
$\epsilon_0=6.0 $ GeV/fm$^3$, so that the total energy in any
rapidity slice remains the same as for the Woods-Saxon initial
density profile. The transverse momentum integrated pion emission
function at midrapidity is shown in Fig. \ref{fig_s_real}. One can
readily note a considerable decrease of the ``averaged" emission
time due to a faster transverse velocity development because of
the initial gradients of density in the whole transverse region
and, therefore, fast cooling. Also we checked the validity of the
Cooper-Frye prescription for spectra at two assumed freeze-out
hypersurfaces corresponding to the isotherms $T = 75 $ MeV and $T
= 160 $ MeV (see Fig. \ref{fig_freezeout_real}), here the  value
of the freeze-out temperature, $T_{f}= 75 $ MeV, is chosen because
it  is near temperatures corresponding to a maximum of the
emission function integrated over transverse momenta.
 We find that the
effective temperature of the spectrum approximately coincides with
the one calculated according to CFp at hypersurfaces $T= 75 $ MeV
for $p_{T}\lesssim 0.5 $ GeV and $T=160 $ MeV for $p_{T}\gtrsim 1$
GeV respectively (see Fig. \ref{fig_spectra_real}).  The spectrum
calculated with HKM is concave, as shown in Ref. \cite{Hama2}, and
demonstrates that the utilization of a simple
$\textbf{p}$-independent isothermal CFp for modeling freeze-out
for such systems could seriously underestimate the effective
temperature of the momentum spectra.

\begin{figure}[h!]
    \centering
        \includegraphics[scale=0.5]{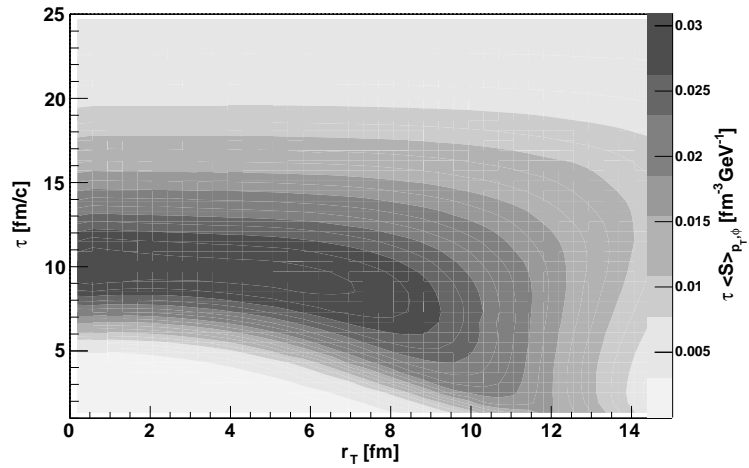}
    \caption{Same as in Fig. \ref{fig_s_real} but with
    an initial nonzero transverse flow added.}
    \label{fig_s_real_flow}
\end{figure}

 Next, in addition to the Gaussian energy-density
profile, we consider also a (pre-equilibrium) nonzero initial flow
in the transverse direction. We set
\begin{equation}
    v_T=\tanh (0.3\cdot{r_T\over R_T}),
    \label{num-3}
\end{equation}
and recalculate the normalization of energy density profile, which
is $\epsilon_0=5.32$ GeV/fm$^3$, so that the total energy of any
rapidity slab of matter in midrapidity remains the same. The
results in the shape of the emission function are shown in Fig.
\ref{fig_s_real_flow}. One can see that the inclusion of the
initial transverse flow leads, as expected, to an even faster
expansion of matter and, thereby, to more reduction of the mean
emission time.

\section{Conclusions}

We have developed the formalism of the hydro-kinetic model
\cite{AkkSin1} intended for a detailed study of the space-time
picture of hadronic emission from rapidly expanding fireballs in
\textit{A}+\textit{A} collisions and, in this way, for an
evaluation of the observed particle spectra and correlations. The
approach developed is consistent with Boltzmann equations and
conservation laws, and accounts also for the opacity effects.

Our analysis and numerical calculations show that the widely used
phenomenological Landau/ Cooper-Frye prescription for the
calculation of pion (or other particles) spectra is too rough if
the freeze-out hypersurface is considered as common for all
momenta of pions. The Cooper-Frye formula, however, could be
applied in a generalized form, accounting for the direct momentum
dependence of the freeze-out hypersurface $\sigma (p) $,
corresponding to the maximum of the emission function $S(t({\bf
r},p),{\bf r},p)$ at fixed momentum $\textbf{p}$ in an appropriate
region of ${\bf r}$. If such a hypersurface $\sigma (p) $ is
found, the conditions of  applicability of the Cooper-Frye formula
{\it for given} $\textbf{p}$ is that the width of the maximum,
which in the simple cases - e.g., for one component system or at
domination of elastic scatterings - is just the relaxation time
(inverse of collision rate), should be smaller than the
corresponding temporal homogeneity length of the distribution
function.

The first approximation within this method is done for fireballs
undergoing 3D azimuthal symmetric Bjorken-type relativistic
expansion. For realistic cross sections, the standard CFp, related
to the fixed isotherm, underestimates the effective temperature of
the observed spectra since, during an ideal hydrodynamic
expansion, the effective temperature of hadronic transverse
spectra is typically decreasing, but particles with relatively
high momenta can escape from the system at an early hot stage. The
latter process is accounted for by the proposed method.

The recent HBT data at RHIC energies restrict a possible maximal
value of the system lifetime, that is a serious problem for
hydro-kinetic ``hybrid" models \cite{hybrid} of matter evolution
in relativistic heavy ion collisions. We studied the effects of
the type/order of phase transition implemented in the EoS, as well
as of different kinds (Woods-Saxon and Gaussian) of an initial
transverse profile, on the duration of pionic emission. We found
that a realistic EoS, motivated by lattice QCD, produces a faster
transverse expansion, reducing therefore the lifetime of the
system as compared with the results based on the EoS with first
order phase transition. The most serious reduction of the lifetime
is observed for the initial Gaussian density profile, because for
this profile there are initial pressure gradients over the whole
transverse slice of the system already at the  initial moment. It
is worth noting  that the Gaussian-type energy density profile
naturally appears \cite{Hirano-Nara} in the Color Glass Condensate
representation of the initial state of colliding nuclei (for
review see, e.g., Ref. \cite{Iancu}) in high energy collisions.
Also the inclusion of the initial nonzero transverse collective
velocity leads to faster expansion of the matter and, thereby,
provides an additional decrease of the emission time. We studied
the effects of inclusion of such flow in the initial conditions
for hydro-kinetic model calculations and found that a moderate
initial flow  developed at the very early prethermal stage of the
evolution of finite partonic systems into vacuum \cite{Sinyukov}
results in a noticeable decrease of the system lifetime. It could
indicate a way for resolving  the HBT puzzle  in the hydro-kinetic
approach. Further developments of the hydro-kinetic approach and
description of the data will be the subject of a follow-up work.

\section*{Acknowledgments}

We are grateful to M. Laine for providing us with the tabulated
equation of state and T. Hirano for his help and advice in
hydrodynamic calculations at the initial stage of this work. This
work was  supported in part by the  FAPESP (Brazil), under
contract nos. 2004/10619-9 and 2006/55393-3;  the Fundamental
Research State Fund of Ukraine, Agreement No. F25/239-2008; the
Bilateral grant DLR (Germany) - MESU (Ukraine) for the UKR 06/008
Project, Agreement No. M/26-2008; and the Program ``Fundamental
Properties of Physical Systems under Extreme Conditions"  of the
Bureau of the Section of Physics and Astronomy of the NAS of
Ukraine.  The research was carried out within the scope of the ERG
(GDRE): Heavy ions at ultrarelativistic energies - a European
Research Group comprising IN2P3/CNRS, Ecole des Mines de Nantes,
Universit\'e de Nantes, Warsaw  University of Technology, JINR
Dubna, ITEP Moscow, and Bogolyubov Institute for Theoretical
Physics, NAS of Ukraine.

\newpage

\newpage

\newpage

\newpage

\newpage

\newpage

\newpage

\newpage

\newpage

\newpage

\newpage

\newpage

\newpage

\newpage

\newpage

\end{document}